\documentclass[twocolumn,aps,superscriptaddress,prb]{revtex4-1}
\setcitestyle{numbers,square}
\usepackage{amsmath,amssymb}
\usepackage{tabularx}
\usepackage{graphicx}
\usepackage{bmpsize}
\usepackage{bm}
\usepackage{color}
\usepackage{amssymb}
\usepackage[version=3]{mhchem}
\usepackage{newtxmath}
\usepackage{braket}
\usepackage[normalem]{ulem} 
\DeclareMathAlphabet{\mathpzc}{OT1}{pzc}{m}{it}
\newcommand{\nn}{\nonumber}
\def\diff{\mathrm d}

\def\i{\mathrm{i}}
\def\e{\mathrm{e}}

\begin{document}
\title{
{Multi-scale lattice relaxation in general twisted trilayer graphenes}
}
\author{Naoto Nakatsuji}
\affiliation{Department of Physics, Osaka University, Toyonaka, Osaka 560-0043, Japan}
\author{Takuto Kawakami}
\affiliation{Department of Physics, Osaka University, Toyonaka, Osaka 560-0043, Japan}
\author{Mikito Koshino}
\affiliation{Department of Physics, Osaka University, Toyonaka, Osaka 560-0043, Japan}
\date{\today}

\begin{abstract}
{
We present comprehensive theoretical studies on the lattice relaxation and the electronic structures in general non-symemtric twisted trilayer graphenes. By using an effective continuum model, we show that the relaxed lattice structure forms a patchwork of moir\'e-of-moir\'e domains, where a moir\'e pattern given by layer 1 and 2 and another pattern given by layer 2 and 3 become locally commensurate. The atomic configuration inside the domain exhibits a distinct contrast between chiral and alternating stacks,
which are determined by the relative signs of the two twist angles.
In the chiral case, the electronic band calculation reveals a wide energy window ($>$ 50 meV) with low density of states, featuring sparsely distributed highly one-dimensional electron bands. These one-dimensional states exhibit a sharp localization at the boundaries between super-moiré domains, and they are identified as a topological boundary state  between distinct Chern insulators.
The alternating trilayer exhibits a coexistence of the flat bands and a monolayer-like Dirac cone, and it is attributed to the formation of moir\'e-of-moir\'e domains equivalent to the mirror-symmetric twisted trilayer graphene.
}
%
\end{abstract}

\maketitle
\section{introduction}
\label{sec:intro}

Two-dimensional moir\'e materials have been the focus of extensive research in recent years. These systems exhibit a long-range moir\'e pattern resulting from lattice mismatch, which profoundly influences their electronic properties. 
Twisted bilayer graphene (TBG), as the most prominent example of a moir\'e system, exhibits the generation of flat bands due to the moir\'e superlattice effect, leading to a variety of correlated quantum phases
\cite{cao2018_80,cao2018_43,doi:10.1126/science.aav1910,Kerelsky2019,xie2019spectroscopic,jiang2019charge,polshyn2019large,Choi2019,doi:10.1126/science.aaw3780,lu2019superconductors,PhysRevLett.124.076801,doi:10.1126/science.aay5533,chen2020tunable,saito2020independent,zondiner2020cascade,wong2020cascade,stepanov2020untying,arora2020superconductivity,PhysRevLett.127.197701}.


 \begin{figure}[h]
  \begin{center}
    \leavevmode\includegraphics[width=0.9\hsize]{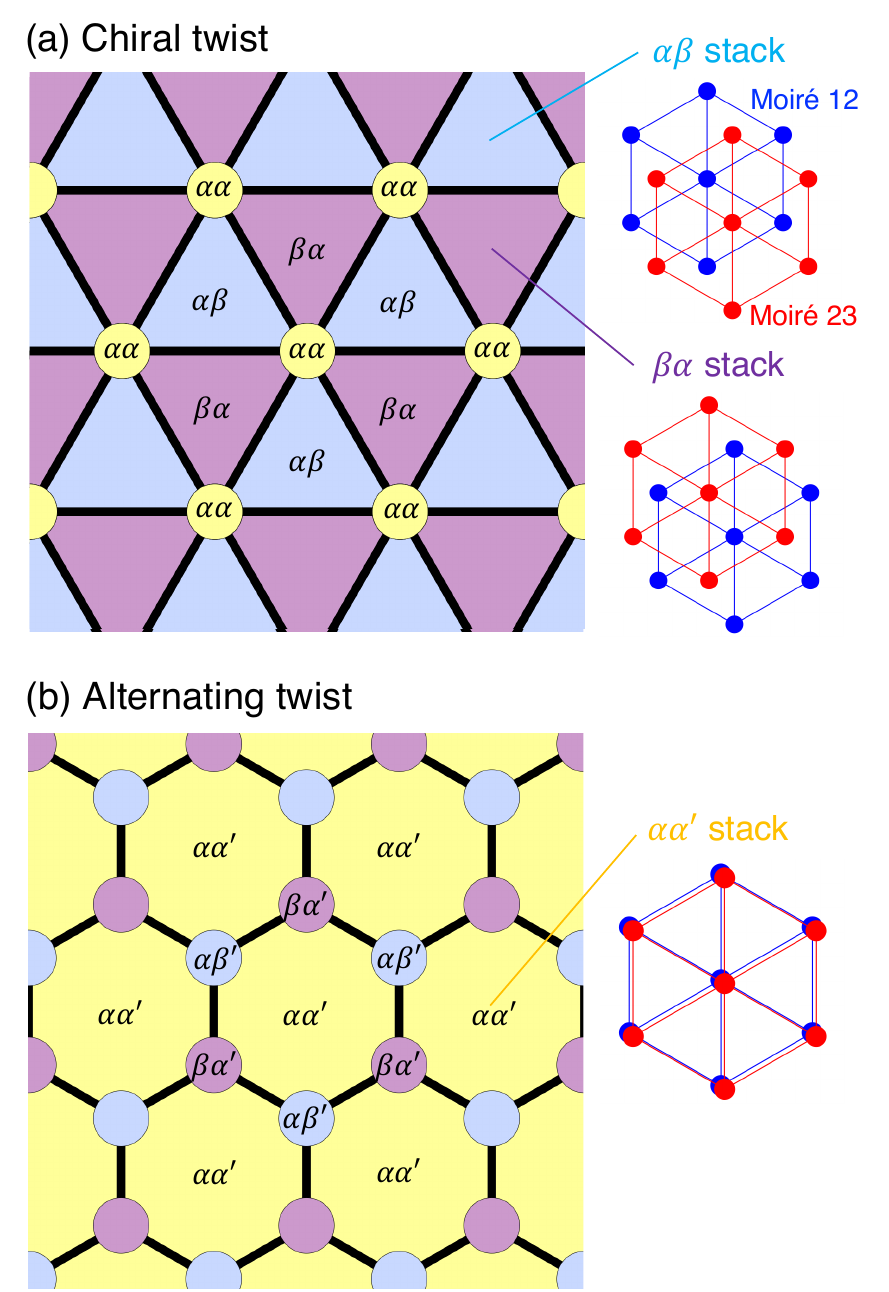}
    \caption{
    {
    Schematic illustration of the moir\'e-of-moir\'e domain structures in (a) chiral TTG and (b) alternating TTG  with close twist angles.
    Right figures represents relative arrangements of two moir\'e patterns within the domains, where blue and red dots indicate AA stacking of moir\'e 12 (between layer 1 and 2) and of moir\'e 23 (between layer 2 and 3), respectively (See also Fig.~\ref{Fig1}).
    }
    }
    \label{Fig0}
  \end{center}
  \end{figure}

In addition to the extensive study of twisted bilayers in the past decade, the scope of investigation has extended to encompass multilayer systems including three or more layers. 
Particular attention has recently been directed towards twisted trilayer graphene (TTG), which consists of three graphene layers arranged in a specific rotational configuration \cite{PhysRevLett.123.026402,li2019electronic,doi:10.1021/acs.nanolett.9b04979,tritsaris2020electronic,PhysRevResearch.2.033357,PhysRevLett.127.026401,PhysRevB.104.035139,PhysRevB.104.L121116,PhysRevB.103.195411,PhysRevB.104.115167,PhysRevResearch.4.L012013,PhysRevX.12.021018,doi:10.1126/science.abn8585,doi:10.1126/science.abg0399,park2021tunable,Cao2021,Kim2022,PhysRevB.101.224107,PhysRevLett.125.116404,lin2020heteromoire,PhysRevLett.127.166802,gao2022symmetry,Ma2022,PhysRevB.105.195422,uri2023superconductivity,PhysRevB.107.125423,popov2023magic,PhysRevB.106.075423,meng2023commensurate}.

\begin{figure*}
  \begin{center}
    \leavevmode\includegraphics[width=0.9\hsize]{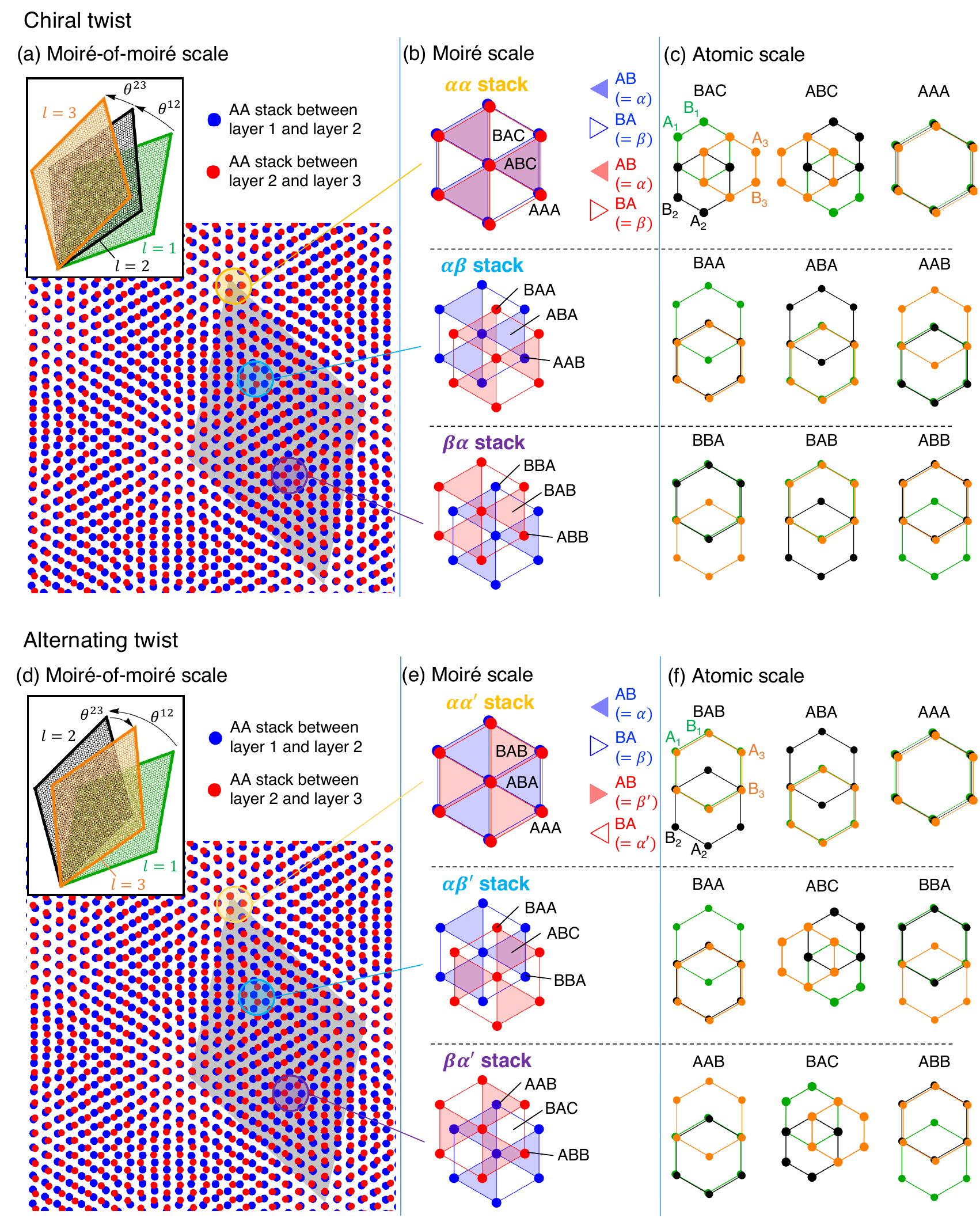}
    \caption{
    (a) Schematics of moir\'e-of-moir\'e pattern of chiral TTG, where blue and red dots represent AA stacking points of moir\'e 12 (between layer 1 and 2) and of moir\'e 23 (between layer 2 and 3), respectively.
    The insert panel illustrates the stacking structure of a chiral TTG, where green, black and orange represent the layer 1, 2 and 3 respectively.
    (b) Local structures of moiré-of-moiré pattern in (a), where circles, filled triangles, and empty triangles indicate AA, AB, and BA stacking of individual moir\'e patterns.
    (c) Local atomic structures at specific points in (b), where $A_{l}$ and $B_{l}$ are the graphene's sublattice in layer $l$.
    {The lower panels [(d), (e) and (f)] are the corresponding figures for the alternate TTG.
    }
    }
    \label{Fig1}
  \end{center}
  \end{figure*}

The system is characterized by twist angles $\theta^{12}$ and $\theta^{23}$ [Fig.~\ref{Fig1}], which represent the relative rotation of layer {2 to 1, and 3 to 2}, respectively.
The special case of $\theta^{12} = -\theta^{23}$ is called the mirror symmetric TTG 
\cite{li2019electronic,doi:10.1021/acs.nanolett.9b04979,tritsaris2020electronic,PhysRevResearch.2.033357,PhysRevLett.127.026401,PhysRevB.104.035139,PhysRevB.104.L121116,PhysRevB.103.195411,PhysRevB.104.115167,PhysRevResearch.4.L012013,PhysRevX.12.021018,park2021tunable,doi:10.1126/science.abg0399,Cao2021,Kim2022,doi:10.1126/science.abn8585}, where layer 1 and layer 3 are aligned precisely, resulting in a single moir\'e periodicity. Recent transport measurements observed correlated insulator phases and robust superconductivity in mirror-symmeric TTGs at a certain magic angle \cite{doi:10.1126/science.abn8585,doi:10.1126/science.abg0399,park2021tunable,Cao2021,Kim2022}.


Beyond the symmetric case, TTG offers a vast parameter space that remains largely unexplored.
In general TTGs with $\theta^{12}\neq -\theta^{23}$, the system has two different moir\'e patterns originating from the interference of layer 1 and 2 and that of layer 2 and 3
\cite{PhysRevLett.123.026402,PhysRevLett.125.116404,lin2020heteromoire,PhysRevLett.127.166802,gao2022symmetry,Ma2022,PhysRevB.105.195422,uri2023superconductivity,PhysRevB.107.125423,popov2023magic,PhysRevB.101.224107,PhysRevB.106.075423,meng2023commensurate}.
These two periodicities are generally incommensurate, giving rise to a quasi-crystalline nature in the system \cite{oka2021fractal,koshino2022topological,uri2023superconductivity}.
When the two moir\'e periods are close but slightly different, in particular, an interference of competing moir\'e structures generate a super-long range moir\'e-of-moir\'e pattern \cite{PhysRevB.101.224107,lin2020heteromoire,PhysRevLett.127.166802,PhysRevLett.125.116404}. Similar situation occurs also in composite multilayer systems consisting of graphene and hexagonal boron nidtride \cite{finney2019tunable,wang2019new,wang2019composite,yang2020situ,andjelkovic2020double,leconte2020commensurate,onodera2020cyclotron,kuiri2021enhanced,shi2021moire,shin2021electron,shin2021stacking,huang2021moir}.
Previous researches investigated the electronic properties of general TTGs with various angle pairs by using several theoretical approaches
\cite{PhysRevLett.123.026402,popov2023magic,Ma2022,PhysRevB.105.195422,uri2023superconductivity,PhysRevLett.127.166802,PhysRevLett.125.116404,PhysRevB.107.125423,PhysRevB.106.075423,meng2023commensurate}.
Recent experimental study also reported superconductivity in some asymmetric TTGs \cite{uri2023superconductivity}.

Generally, twisted moir\'e systems are under a strong influcence of lattice relaxation in the moir\'e scale, which also significantly modifies the electronic properties.
In TBG, for instance, an in-plane lattice distortion forms commensurate AB(Bernal)-stacking domains 
\cite{popov2011commensurate, brown2012twinning,lin2013ac,alden2013strain,uchida2014atomic,van2015relaxation,dai2016twisted,jung2015origin, jain2016structure,nam2017lattice,carr2018relaxation, lin2018shear,yoo2019atomic,guinea2019continuum,koshino2020effective},
and it opens energy gaps in the electronic spectrum to isolate low-energy flat bands \cite{nam2017lattice,carr2018relaxation, koshino2020effective}.
The lattice relaxation occurs also in trilayer moir\'e systems,
where the moir\'e-of-moir\'e period superstructure was observed \cite{doi:10.1126/science.abk1895,li2022symmetry,craig2023local}.
Such a large-scale relaxation was also theoretically simulated for various trilayer systems \cite{PhysRevB.101.224107,shin2021electron,shin2021stacking,PhysRevB.106.075423,meng2023commensurate}.


In this paper, we study the lattice relaxation and 
the electronic band structure in non-symmetric TTGs.
TTG is classified into two groups depending on the relative direction of rotation angles; the cases of $\theta^{12} \cdot \theta^{23} > 0$ and $< 0$, which are refereed to as chiral and alternating TTGs, respectively \cite{Ma2022,PhysRevB.105.195422,uri2023superconductivity}.
{Here we consider chiral and alternating TTGs having various combinations of twist angles $(\theta^{12}, \theta^{23})$.}
We obtain the optimized lattice structure using the effective continuum approach used for TBG \cite{nam2017lattice,PhysRevB.100.075416,PhysRevB.107.115301},
and compute the electronic structure by a continuum band calculation method including the lattice relaxation \cite{koshino2020effective}.

We find that there are two distinct length-scale
relaxations in the moir\'e-of-moir\'e and moir\'e scales, which
give rise to a formation of a patchwork of super-moir\'e domains as schematically shown in Fig.~\ref{Fig0}. 
{
In theses domains, the first moir\'e pattern given by layer 1 and 2 (moir\'e 12) and the second pattern by layer 2 and 3 (moir\'e 23) are deformed to become commensurate. The atomic configuration inside the domain exhibits a distinct contrast between chiral and alternating TTGs: In the chiral case,
the two moir\'e patterns are arranged such that the AA spots of moir\'e 12 and those of moir\'e 23 repel to each other, leading to shifted configurations [Fig.~\ref{Fig0}(a)].
In the alternating case, in contrast, the AA spots attract each other, resulting in a fully overlapped structure equivalent to the  mirror-symmetric TTG  [Fig.~\ref{Fig0}(b)].
The energetic stability of these super-moir\'e domain formations can be explained by considering 
a competition of lattice relaxation in the two moir\'e patterns.
}

In the band calculation, we find that the spectrum
{of the chiral TTG} has an energy window more than 50 meV wide with low density of state, where highly one-dimenisinoal electron bands are sparsely distributed.
The wave function of the one-dimensional bands is sharply localized at the boundary between the super-moir\'e domains.
By calculating the Chern number of the local band structure of the commensurate domain, the one-dimensinal state is shown to be a topological boundary state between distinct Chern insulators.
{On the other hand, the alternating TTG
exhibits a coexistence of the flat bands and a monolayer-like Dirac cone, resembling the energy spectrum of the mirror-symmetric TTG \cite{park2021tunable,doi:10.1126/science.abg0399,Cao2021,doi:10.1126/science.abn8585,Kim2022}. Here the moir\'e-of-moir\'e relaxation significantly reduces the hybridization of the Dirac cone with other states, restoring its highly-dispersive feature.
}




 {
The paper is organized as follows.
In Sec.~\ref{sec_model}, we define the lattice structure of TTG and introduce the continuum method to calculate of the lattice relaxation and the electronic band structure.
In Sec.~\ref{sec_chiral}, we investigate the chiral TTGs.
We obtain the relaxed lattice structure and demonstrate the formation of the moir\'e-of-moir\'e domain pattern in Sec.~\ref{sec_lattice_relaxation_chiral}.
We calculate the band structure including the lattice relaxation in Sec.~\ref{sec_elec_chiral}, where we show the emergence of the one-dimensional boundary states on the domain walls.
In Sec.~\ref{sec_alternating}, we conduct similar analyses for the alternating TTGs.
}



  \begin{figure*}
  \begin{center}
    \leavevmode\includegraphics[width=1. \hsize]{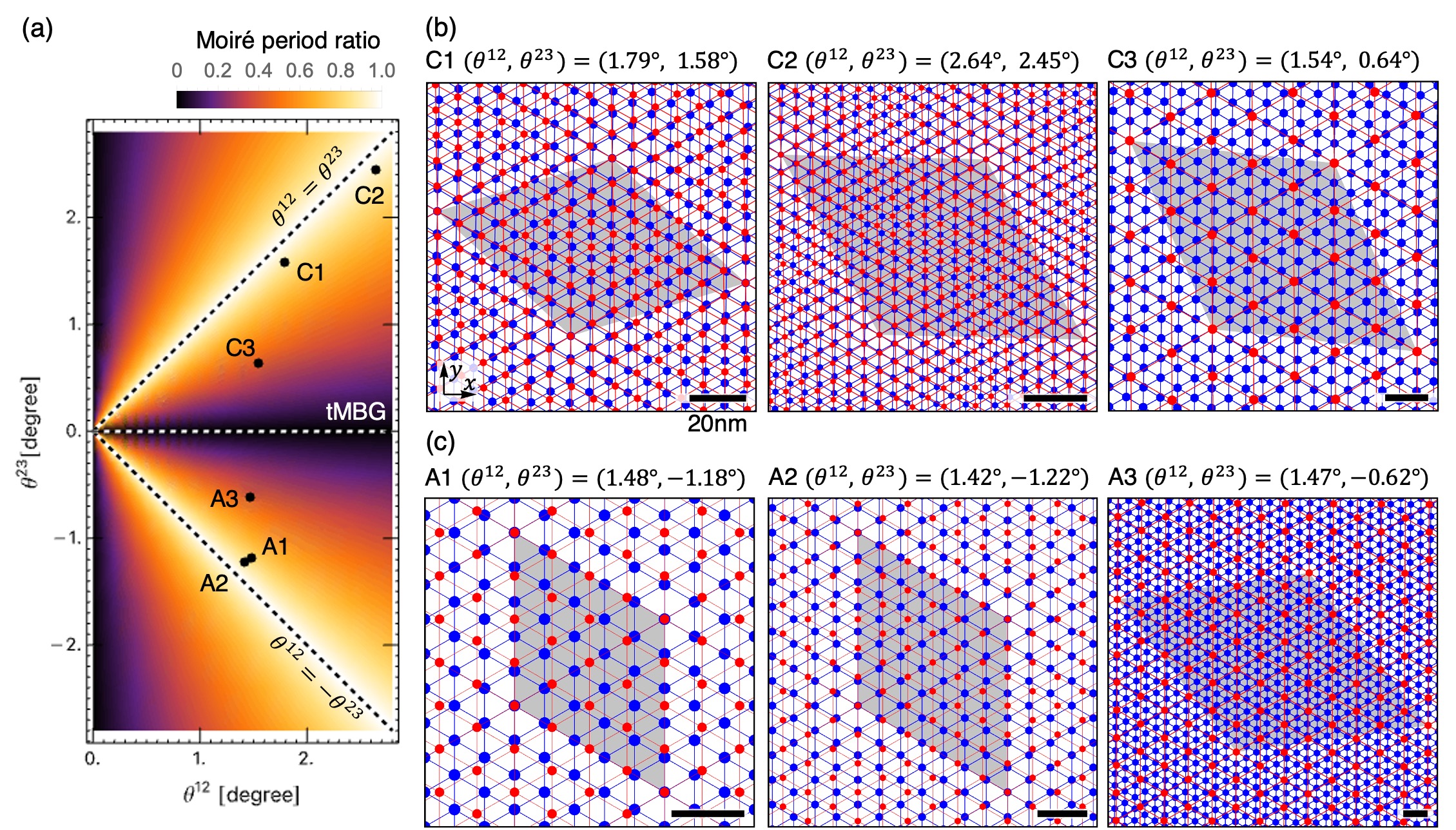}
    \caption{
{
(a) Two-dimensional map of $(\theta^{12},\theta^{23})$ of TTGs considered in this paper.
The color code represents the ratio of the two moir\'e periods, $\min\left(L^{12}/L^{23},L^{23}/L^{12}\right)$.
Diagonal dashed lines indicate $\theta^{12}=\pm\theta^{23}$, and a horizontal dashed line represents twisted monolayer-bilayer graphene (tMBG). 
 (Right) Moir\'e-of-moir\'e patterns without lattice relaxation of (b) chiral TTGs (C1, C2 and C3) and (c) alternating TTGs (A1, A2 and A3). 
 Blue and red dots indicate the AA spot of moir\'e 12 (between layer 1 and 2) and moir\'e 23 (between layer 2 and 3) respectively, and gray area represents the moir\'e-of-moir\'e unit cell. 
 All scale bars indicate 20 nm.
    }
    }
    \label{Fig3}
  \end{center}
  \end{figure*}

\section{Model}
\label{sec_model}

\subsection{Geometry of TTG}
\label{sec_geometry}
    {
We define a TTG by stacking three graphene layers labeled by $l=1,2$ and 3, with relative twist angles $\theta^{12}$ (layer 1 to 2) and $\theta^{23}$ (layer 2 to 3).
{
The configuration is schematically depicted in Fig.~\ref{Fig1}(a) and (b), 
for the chiral case
($\theta^{12}\cdot\theta^{23} > 0$)
and the alternating case ($\theta^{12}\cdot\theta^{23} < 0$), respectively.
}
The primitive lattice vectors of layer $l$ are defined by $\bm{a}_{i}^{(l)}=R(\theta^{(l)})\bm{a}_i$ where $\bm{a}_1=a(1,0)$ and $\bm{a}_2=a(1/2,\sqrt{3}/2)$ are the lattice vectors of unrotated monolayer graphene, $a=0.246$ nm is the graphene’s lattice constant. $R$ is the rotation matrix, and
$\theta^{(l)}$ is the absolute twist angle of layer $l$ given by $\theta^{(1)}=-\theta^{12}$, $\theta^{(2)}=0$ and $\theta^{(3)}=\theta^{23}$.
Accordingly, the primitive reciprocal lattice vectors become $\bm{b}_{i}^{(l)}=R(\theta^{(l)})\bm{b}_i$ where $\bm{b}_1=(2\pi/a)(1,-1/\sqrt{3})$ and $\bm{b}_2=(2\pi/a)(0,2/\sqrt{3})$ are the reciprocal lattice vectors without rotation.
The Dirac points of graphene layer $l$ are intrinsically located at the corners of Brillouin zone (BZ), $K_{\xi}^{(l)}=-\xi\left(2\bm{b}^{(l)}_{1}+\bm{b}^{(l)}_{2}\right)/3$ where $\xi = \pm 1$ is the valley index.
}


{In this paper, we consider TTGs with small twist angles ($|\theta^{12}|, |\theta^{23}| \lesssim 10^{\circ}$).}
Then the system is governed by two competing moir\'e patterns, one from the layer 1 and 2 and the other from layer 2 and 3.
The reciprocal lattice vectors for these moir\'e patterns are given by $\bm{G}_{i}^{ll'}=\bm{b}_{i}^{(l)}-\bm{b}_{i}^{(l')}$ where $(l,l') = (1,2)$ or $(2,3)$. 
{
The moir\'e lattice vectors can be obtained from $\bm{G}_{i}^{ll'}\cdot\bm{L}_{j}^{ll'}=2\pi\delta_{ij}$, and explicitly written as
\begin{align}
  & \bm{L}^{12}_{1} = 
  \frac{a}{2\sin{(\theta^{12}/2)}}
  R(-\theta^{12}/2)
  \begin{pmatrix}
      0 \\ -1
  \end{pmatrix}
 \notag\\
&  \bm{L}^{23}_{1} = 
  \frac{a}{2\sin{(\theta^{23}/2)}}
  R(+\theta^{23}/2)
  \begin{pmatrix}
      0 \\ -1
  \end{pmatrix}, \label{eq_L_moire}
\end{align}
and $\bm{L}^{ll'}_{2}=R(60^\circ)\bm{L}^{ll'}_{1}$.}
The moir\'e lattice constant is given by $L^{ll'}=|\bm{L}^{ll'}_{1}|=|\bm{L}^{ll'}_{2}|=a/|2\sin{(\theta^{ll'}/2)}|$.

{When absolute twist angles are close $(|\theta^{12}|\approx |\theta^{23}|)$,} an interference between the two moir\'e patterns gives rise to a higher order structure called a moir\'e-of-moir\'e pattern as shown in Fig.~\ref{Fig1}.
{Here the upper and lower rows correspond to the chiral and alternating structures, respectively.
For the chiral twist, the left panel [Fig.~\ref{Fig1}(a)] illustrates the overlapped moir\'e patterns where blue and red dots represent the AA spots of moir\'e 12 and 23, respectively.
The local structure can be viewed as a pair of non-twisted moir\'e superlattices with a relative translation, as illustrated in Fig.~\ref{Fig1}(b).
Here shaded and empty triangles represent AB, and BA stacking regions of individual moir\'e patterns, respectively.
By defining AB and BA points (the centers of triangles) by $\alpha$ and $\beta$, respectively,
the local stacking configuration of the two moir\'e patterns is 
labeled by $\alpha\alpha$, $\alpha\beta$ and $\beta\alpha$.
Figure \ref{Fig1}(c) depicts the local structure in the atomic scale.
Here $A_l$ and $B_l$ represent the graphene's sublattice 
in layer $l$. We define the sublattice $C_l$ as the center of the hexagon in the honeycomb lattice. For instance, BAC-stacking represents 
$B_1$, $A_2$ and $C_3$ are vertically aligned.
}

{
The lower panels [Figs.~\ref{Fig1}(d), (e) and (f)] are the corresponding figures for the alternate twist.
The key difference from the chiral case lies in the 180$^\circ$ rotation of the moir\'e 23 (red lattice) due to the opposing sign of $\theta^{23}$.
This results in the flipping of the positions of AB and BA.
Consequently, the local atomic structure (shown in the rightmost panels) differs between the chiral and alternating structures, even though the relative arrangement of AA spots is identical.
We define AB and BA points in the inverted moir\'e 23 pattern by $\beta'$ and $\alpha'$, respectively,
and label the local structure in the alternating TTG
by $\alpha\alpha'$, $\alpha\beta'$ and $\beta\alpha'$,
as in Fig.~\ref{Fig1}(e).
}

\subsection{Commensurate TTGs}

{
Generally the two moir\'e patterns in a TTG are not commensurate,
and the spatial period of moir\'e-of-moir\'e pattern is infinite.
However, there are special angle sets $(\theta^{12},\theta^{23})$ where the two patterns happen to have a finite common period.
In such a case, we can express the moir\'e-of-moir\'e  primitive lattice vectors $\bm{L}_{1}$ and $\bm{L}_{2}$
in terms of integers $n, m, n'$ and $m'$ as
\begin{align}\label{eq_comensurate_condition}
&        \bm{L}_{1} = n\bm{L}^{12}_{1}+m\bm{L}^{12}_{2} = n'\bm{L}^{23}_{1}+m'\bm{L}^{23}_{2},
\notag\\
&  \bm{L}_{2} = R(60^\circ)\bm{L}_{1}.
    \end{align}
The moir\'e-of-moir\'e reciprocal lattice vectors are given by the condition $\bm{G}_{i}\cdot\bm{L}_{j}=2\pi\delta_{ij}$.
{
The corresponding twist angles are obtained 
by solving Eqs.~\eqref{eq_L_moire} and \eqref{eq_comensurate_condition} for variables $\theta^{12}$ and $\theta^{23}$,
as}
    \begin{align}\label{eq_angle_formulas}
   \theta^{12} = \theta(n,m,n',m'), \quad 
   \theta^{23} = - \theta(n',m',n,m),
%
    \end{align}
where 
\begin{align}\label{eq_angle_formulas_2}
 &\theta(n,m,n',m') =\notag\\
 & \quad 2 \tan^{-1}\frac{\sqrt{3}\left\{m \left(2n'+m'\right)-\left(2n+m\right)m'\right\}}{\left(2n+m\right)\left(2n'+m'\right)+3mm'+\left(2n'+m'\right)^{2}+3m'^{2}}.
\end{align}
The spatial period of the super-moir\'e pattern is given by 
$L = L^{12}\sqrt{n^{2}+m^{2}+nm}= L^{23}\sqrt{n'^{2}+m'^{2}+n'm'}$.

{
In alternating TTGs with $\theta^{12}\approx -\theta^{23}$,
the relative angle between two moir\'e lattice vectors nearly vanishes, resulting in an extremely large commensurate moiré-of-moiré unit cell.
To treat such cases, we neglect the tiny misorientation of the moir\'e lattice vectors $\bm{L}_{j}^{12}$ and $\bm{L}_{j}^{23}$, while retaining their norms. 
In this approximation, the moir\'e-of-moir\'e commensurate period is expressed as
\begin{align}
\label{eq_comensurate_condition_approx} 
\bm{L}_{1} = n\bm{L}^{12}_{1} = n'\bm{L}^{23}_{1},
\quad \bm{L}_{2} = R(60^\circ)\bm{L}_{1},
\end{align}
instead of Eq.\eqref{eq_comensurate_condition}.
Note that Eq.~\eqref{eq_angle_formulas}
does not apply to this approximate commensurate structure.}

\begin{table}
    \centering
    \[
\begin{array}{|c|c|c|c|}
\hline
& (\theta^{12},\theta^{23}) & (n,m,n',m')& L^{12}/L^{23}\\
\hline 
{\rm C1} & (1.79^\circ,1.58^\circ) &  (2,7,2,6) & 0.88 \\
{\rm C2} & (2.64^\circ,2.45^\circ) & (7,7,7,6) & 0.93 \\
{\rm C3} & (1.54^\circ,0.64^\circ) &  (7,5,3,2) & 0.42 \\
\hline
{\rm A1} & (1.48^\circ,-1.18^\circ) &  (5, 0, -4, 0)^* & 0.80 \\
{\rm A2} & (1.42^\circ,-1.22^\circ) & (7, 0, -6, 0)^* & 0.86 \\
{\rm A3} & (1.47^\circ,-0.62^\circ) &  (7, 12, -3, -5) & 0.42 \\
\hline
\end{array}
    \]
    \caption{
    {Definition of commensurate chiral TTGs (C1, C2, C3) and commensurate alternating TTGs (A1, A2, A3) considered in this paper. The asterisk (*) symbol for A1 and A2 indicates the use of the approximation of Eq.~\eqref{eq_comensurate_condition_approx} to obtain the commensurate structures.}
    }
    \label{tab:C123}
\end{table}

{
In this paper, we consider commensurate chiral TTGs, C1, C2 and C3, and commensurate alternating TTGs, A1, A2 and A3, 
defined in Table \ref{tab:C123}.
We employ the exact commensurate formulas Eqs.~\eqref{eq_comensurate_condition} and \eqref{eq_angle_formulas}
for C1, C2, C3, and A3,
while we utilize the approximate formula, Eq.~\eqref{eq_comensurate_condition_approx} for A1 and A2.
Figure \ref{Fig3}(a) maps
$(\theta^{12},\theta^{23})$ of these systems in two-dimensional space,
where the color code represents the ratio of the two moir\'e periods, $\min\left(L^{12}/L^{23},L^{23}/L^{12}\right)$.
The moir\'e-of-moir\'e structures of these TTGs without lattice relaxation are illustrated in Fig.~\ref{Fig3}(b) and (c), respectively.
}

{
    We show the schematics of Brillouin zone (BZ) of a chiral TTG in Fig.~\ref{Fig4}. Here green, black and orange hexagons represent the first BZ of layer 1, 2, and 3, respectively.
    Blue and red hexagons represent the BZ for the first moir\'e patterns given by $l=1,2$ and the second pattern given by $l=2,3$, respectively.
    Finally, the gray heaxagon is the BZ of the moir\'e-of-moir\'e pattern, where
    we label the corner points by $\kappa$ and $\kappa'$, the midpoint of a side by $\mu$ and the center by $\gamma$. 
    }

  \begin{figure}
  \begin{center}
    \leavevmode\includegraphics[width=0.8 \hsize]{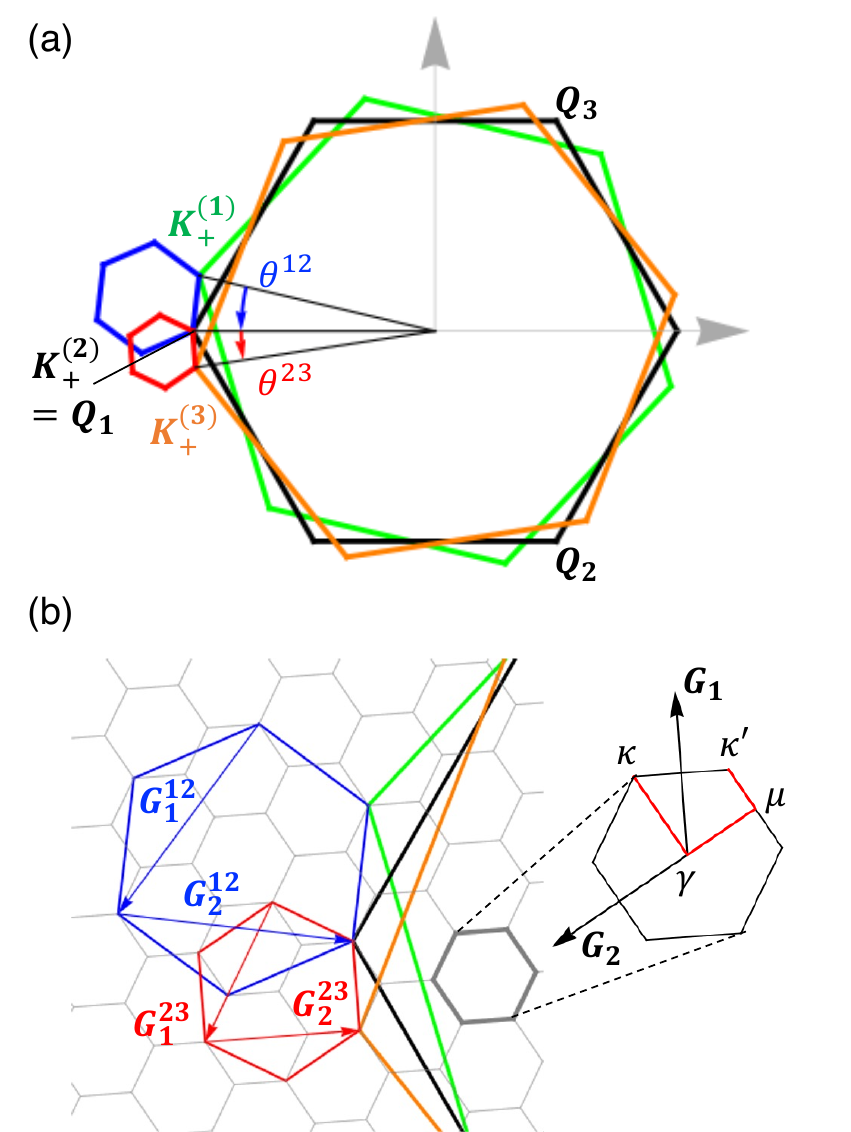}
    \caption{
    Brillouin zone of chiral TTG. Green, black and orange hexagons represent the first Brillouin zone of graphene layer 1, 2, and 3, respectively.
    Blue and red hexagons represent the BZ for the moir\'e patterns given by $l=1,2$ and that by $l=2,3$, respectively.
    Gray heaxagon is the BZ of the moir\'e-of-moir\'e pattern.
    }
    \label{Fig4}
  \end{center}
  \end{figure}
  
\subsection{Continuum method for multi-scale lattice relaxation}
\label{sec:mslr}

{
    We adopt a continuum approximation 
\cite{nam2017lattice,PhysRevB.100.075416,PhysRevB.107.115301} 
    to describe the lattice relaxation on TTG.
   Let $\bm{s}^{(l)}(\bm{R}_X)$ be the displacement vector of sublattice $X = A$, or $B$
   at a two-dimenisonal position $\bm{R}_X$ of layer $l=1,2,3$.
    Here we consider a long-rage lattice relaxation which has much longer scales than graphene's lattice constant.  The desplacement vectors can then be 
    expressed by continuous functions in real space as
    $\bm{s}^{(l)}(\bm{R}_A)=\bm{s}^{(l)}(\bm{R}_B)=\bm{s}^{(l)}\left(\bm{r}\right)$.
    We ignore the out-of-plane component of the displacement vector in this model, as it does not much contribute to the commensurate domain formation. 
    The optimized lattice structure can be obtained by minimizing the total energy $U=U_{E}+U_{B}^{12}+U_{B}^{23}$,
    where $U_{E}$ is the elastic energy and $U_{B}^{ll'}$ is the interlayer binding energy between layers $l$ and $l'$.
    We assume that $U_{B}^{12}$ and $U_{B}^{23}$ are given by the interlayer interaction energy of the twisted bilayer graphene \cite{nam2017lattice},
    and neglect a remote interaction between layer 1 and 3. 
    The $U_{E}$ and $U_{B}^{ll'}$ can be expressed as funtionals of the displacement field $\bm{s}^{(l)}\left(\bm{r}\right)$.
    We solve the Euler-Lagrange equation to obtain the optimized $\bm{s}^{(l)}\left(\bm{r}\right)$ self-consistently.
}

{    
 The elastic energy of strained TTG is written in a standard form \cite{PhysRevB.65.235412,PhysRevB.90.115152} as
    \begin{align}\label{eq:elastic}
        U_E=\sum_{l=1}^{3}\frac{1}{2}\int\left[\left(\mu+\lambda\right)\left(s_{xx}^{(l)}+s_{yy}^{(l)}\right)^{2} \right.\notag \\
           \left. +\mu\left\{\left(s_{xx}^{(l)}-s_{yy}^{(l)}\right)^{2}+4\left(s_{xy}^{(l)}\right)^{2}\right\}\right]\diff^2\bm{r},
    \end{align}
    where $\lambda=3.25$~eV/$\AA^{2}$ and $\mu=9.57$~eV/$\AA^{2}$ are graphene's Lam\'e factors\cite{PhysRevLett.102.046808,jung2015origin}, and $s_{ij}^{(l)}=(\partial_{i}s_{j}^{(l)}+\partial_{j}s_{i}^{(l)})/2$ is the strain tensor.
    The interlayer binding energy of adjacent layers $(l,l')=(1,2), (2,3)$
    is given by \cite{nam2017lattice}
     \begin{align}\label{eq:binding}
     U_{B}^{ll'}&=\int\diff^{2}\bm{r} \sum_{j=1}^{3}2V_{0}\cos\left[\bm{G}_{j}^{ll'}\cdot\bm{r}+\bm{b}_{j}\cdot\left(\bm{s}^{(l')}-\bm{s}^{(l)}\right)\right], 
    \end{align}
    where $\bm{b}_{3}=-\bm{b}_{1}-\bm{b}_{2}$, $\bm{G}_{3}^{ll'}=-\bm{G}_{1}^{ll'}-\bm{G}_{2}^{ll'}$. 
    We take $V_{0}= 0.160$~eV/nm$^{2}$\cite{PhysRevB.84.045404,lebedeva2011interlayer}.
}

{
 We introduce 
\begin{align}\label{eq:transform}
        \bm{w} &= \bm{s}^{(1)}+\bm{s}^{(2)}+\bm{s}^{(3)} \nn \\ 
        \bm{u} &= \bm{s}^{(1)}-2\bm{s}^{(2)}+\bm{s}^{(3)} \nn \\
        \bm{v} &= \bm{s}^{(1)}-\bm{s}^{(3)},
    \end{align}
    and rewrite $U$ as a functional of $\bm{w}, \bm{u}$ and $\bm{v}$.
Here $\bm{w}$ represents an overall translation of three layers, while $\bm{u}$ and $\bm{v}$ are relative slidings
which are mirror-even and odd, respectively, with respect to the middle layer.
In the subsequent analysis, we fix $\bm{w}$ to zero and focus solely on $\bm{u}$ and $\bm{v}$, as $\bm{w}$ does not alter the interlayer registration and therefore does not impact the formation of moir\'e domains.
The Euler-Lagrange equation is written as    \begin{align}\label{eq_Euler_Lagrange}
        \hat{K}\bm{u} + 6V_{0}\sum_{j=1}^{3} \left\{\sin\left[\bm{G}_{j}^{12}\cdot\bm{r}-\bm{b}_{j}\cdot\left(\bm{u}+\bm{v}\right)/2\right]\right. \notag \\
                                      \left.+\sin\left[\bm{G}_{j}^{23}\cdot\bm{r}+\bm{b}_{j}\cdot\left(\bm{u}-\bm{v}\right)/2\right]\right\}\bm{b}_{j}=0 \\
        \hat{K}\bm{v} + 2V_{0}\sum_{j=1}^{3} \left\{\sin\left[\bm{G}_{j}^{12}\cdot\bm{r}-\bm{b}_{j}\cdot\left(\bm{u}+\bm{v}\right)/2\right]\right. \notag \\
                                      \left.-\sin\left[\bm{G}_{j}^{23}\cdot\bm{r}+\bm{b}_{j}\cdot\left(\bm{u}-\bm{v}\right)/2\right]\right\}\bm{b}_{j}=0,
    \end{align}
    where
    \begin{align}\label{eq_K_define}
        \hat{K}=
        \left(
		\begin{array} {cc}
		\left(\lambda+2\mu\right)\partial_{x}^{2}+\mu \partial_{y}^{2} & \left(\lambda+\mu\right)\partial_{x}\partial_{y} \\
		\left(\lambda+\mu\right)\partial_{x}\partial_{y} & \left(\lambda+2\mu\right)\partial_{y}^{2}+\mu \partial_{x}^{2}
		\end{array}
	\right).
    \end{align}
 }

 {
We assume $\bm{s}^{(l)}$'s (so $\bm{u}$ and $\bm{v}$) are periodic in the original moir\'e-of-moir\'e period, and define the Fourier components as
 \begin{align}\label{eq:Fourier_transform_uv}
  \bm{u}\left(\bm{r}\right) =   \sum_{\bm{G}}\bm{u}_{\bm{G}}\e^{\i\bm{G}\cdot\bm{r}},
\quad
\bm{v}\left(\bm{r}\right) =   \sum_{\bm{G}}\bm{v}_{\bm{G}}\e^{\i\bm{G}\cdot\bm{r}}, 
\end{align}
where $\bm{G}=m_{1}\bm{G}_{1}+m_{2}\bm{G}_{2}$ are the moir\'e-of-moir\'e reciprocal lattice vectors.
We also introduce $f_{\bm{G},j}^{ll'}$ by
\begin{align}
&\sin\left[\bm{G}_{j}^{12}\cdot\bm{r}-\bm{b}_{j}\cdot(\bm{u}+\bm{v})/2\right]
=  \sum_{\bm{G}}f_{\bm{G},j}^{12}\e^{\i\bm{G}\cdot\bm{r}},
\notag\\
&\sin\left[\bm{G}_{j}^{23}\cdot\bm{r}+
\bm{b}_{j}\cdot(\bm{u}-\bm{v})/2\right]
=  \sum_{\bm{G}}f_{\bm{G},j}^{23}\e^{\i\bm{G}\cdot\bm{r}}.
\label{eq:Fourier_transform}
\end{align}
%
%
%
Eq.~\eqref{eq_Euler_Lagrange} is then written as
    \begin{align}\label{eq:static_sc}
        \bm{u}_{\bm{G}} &= -6V_{0}\sum_{j=1}^{3}\left(f_{\bm{G},j}^{12}+f_{\bm{G},j}^{23}\right)\hat{K}_{\bm{G}}^{-1}\bm{b}_{j}, \notag \\
        \bm{v}_{\bm{G}} &= -2V_{0}\sum_{j=1}^{3}\left(f_{\bm{G},j}^{12}-f_{\bm{G},j}^{23}\right)\hat{K}_{\bm{G}}^{-1}\bm{b}_{j},
    \end{align}
    where
    \begin{align} \label{eq:def_K}
        \hat{K}_{\bm{G}} =
	    \left(
			\begin{array} {cc}
			\left(\lambda+2\mu\right)G_{x}^{2}+\mu G_{y}^{2} & \left(\lambda+\mu\right)G_{x}G_{y} \\
			\left(\lambda+\mu\right)G_{x}G_{y} & \left(\lambda+2\mu\right)G_{y}^{2}+\mu G_{x}^{2}
			\end{array}
		\right).
    \end{align}
%
}

{
We obtain the optimized $\bm{u}_{\bm{G}}$ and $\bm{v}_{\bm{G}}$ by 
solving Eqs.~\eqref{eq:Fourier_transform} and \eqref{eq:static_sc} in an iterative manner.
In the calculation, we only consider a finite number of the Fourier components in $|\bm{G}|<3\max \left(|n|,|m|,|n'|,|m'|\right)$, which are sufficient to describe the lattice relaxation in the systems considered.
It should be noted that the components of $\bm{G}=0$ cannot be 
determined by this scheme, since $\hat{K}_{\bm{G}}$ becomes 0 in Eq.~\eqref{eq:static_sc}.
Here we treat $\bm{s}^{(l)}_{\bm{G}=0}$ as parameters, and perform the above iteration for different parameter choices. We finally choose the solution having the lowest total energy. 
The dependence on $\bm{G}=0$ component arises because the moir\'e-of-moir\'e structure depends on a relative translation of the two moir\'e patterns, and hence it cannot be eliminated by a shift of the origin unlike twisted bilayer graphene. Practically, it is sufficient to consider only the lateral sliding of layer 3 with other two layers fixed.
}


    
\subsection{Continuum Hmiltonian with lattice relaxation}
\label{subsec:CM}

{
 We compute the band structure of the TTGs by
 using an electronic continuum model \cite{lopes2007graphene,bistritzer2011moirepnas,kindermann2011local,PhysRevB.86.155449,moon2013opticalabsorption,koshino2015interlayer}
 that incorporates lattice relaxation \cite{koshino2020effective}.
  The effective Hamiltonian for valley $\xi$ is written as 
    \begin{align} \label{eq_Hamiltonian}
        {H}^{(\xi)} =
	    \left(
			\begin{array} {ccc}
		      H_{1}\left(\bm{k}\right) & U_{21}^{\dagger} & \\
		      U_{21} & H_{2}\left(\bm{k}\right) & U_{32}^{\dagger}\\
                 & U_{32} & H_{3}\left(\bm{k}\right),
			\end{array}
		\right).
    \end{align}
    The matrix works on
    a six-component wave function  $(\psi_{A}^{(1)}, \psi_{B}^{(1)}, \psi_{A}^{(2)}, \psi_{B}^{(2)}, \psi_{A}^{(3)}, \psi_{B}^{(3)})$, 
    where $\psi_{X}^{(l)}$ represents the envelope function of sublattice $X(=A,B)$ on layer $l(=1,2,3)$. 
    The $H_l(\bm{k})$ is the $2\times 2$ Hamiltonian of monolayer graphene and $U_{ll'}$ is the interlayer coupling matrix, in the presence of the lattice distortion.
    The $H_l(\bm{k})$ is given by
    \begin{flalign} \label{eq:monolayerGrapheneHamiltonian}
    H_{l}(\bm{k})=-\hbar v\left[R\left(\theta^{(l)}\right)^{-1} \left(\bm{k}-\bm{K}_{\xi}^{(l)}+\frac{e}{\hbar}\bm{A}^{(l)}\right)\right]\cdot \bm{\sigma},
    \end{flalign}
    where $v$ is the graphene's band velocity, $\bm{\sigma} = \left(\xi \sigma_{x}, \sigma_{y}\right)$ and $\sigma_{x}$, $\sigma_{y}$ are the Pauli matrices in the sublattice space $\left(A, B\right)$.    
    We take $\hbar v/a = 2.14$ eV \cite{PhysRevX.8.031087}. 
    The $\bm{A}^{(l)}$ is the strain-induced vector potential that is given by \cite{PhysRevB.65.235412,PhysRevLett.103.046801,Guinea2010}
    \begin{eqnarray}\label{eq_A1}
        \bm{A}^{(l)} &=& 
	\xi \frac{3}{4}\frac{\beta\gamma_{0}}{ev}
        \begin{pmatrix}
            s_{xx}^{(l)}-s_{yy}^{(l)}\\
            - 2 s_{xy}^{(l)}
        \end{pmatrix},
    \end{eqnarray} 
    where $\gamma_{0}=2.7$ eV is the nearest neighbor transfer energy of intrinsic graphene and $\beta \approx 3.14$.
    }

{
    The interlayer coupling matrix $U_{21}$ and $U_{32}$ are given by
    \begin{align} \label{eq:Moire_interaction}
	U_{l'l}= \sum_{j=1}^{3} U_{j} \e^{\i \xi\bm{\bm{\delta k}}_{j}^{(ll')}\cdot\bm{r} + \i\bm{Q}_{j}\cdot\left(\bm{s}^{(l')}-\bm{s}^{(l)}\right)} 
    \end{align}
where we defined
\begin{align}
\label{eq_dk}
&\delta\bm{k}_{1}^{ll'}=\bm{0},~~~\delta\bm{k}_{2}^{ll'}=\xi\bm{G}_{1}^{ll'},~~~\delta\bm{k}_{3}^{ll'}=\xi\left(\bm{G}_{1}^{ll'}+\bm{G}_{2}^{ll'}\right),
\\
\label{eq_qj}
&\bm{Q}_{1}=\bm{K}_{\xi},~~~\bm{Q}_{2}=\bm{K}_{\xi}+\xi\bm{b}_{1},~~~\bm{Q}_{3}=\bm{K}_{\xi}+\xi\left(\bm{b}_{1}+\bm{b}_{2}\right),
\end{align}
and
    \begin{align} \label{eq_u_mat}
	&U_{1}=\left(
		\begin{array}{cc}
		u & u' \\
		u' & u 
		\end{array}
	\right), \quad
	U_{2}=\left(
		\begin{array}{cc}
		  u & u'\omega^{-\xi} \\
		u'\omega^{+\xi} & u
		\end{array}
	   \right), 
        \notag\\
	&U_{3}=\left(
		\begin{array}{cc}
		u & u'\omega^{+\xi} \\
		u'\omega^{-\xi} & u
		\end{array}
	\right). 
    \end{align}
The parameters $u=79.7$ meV and $u'=95.7$ meV are interlayer coupling strength between AA/BB and AB/BA stack region, respectively\cite{PhysRevX.8.031087, PhysRevB.101.195425}.
In the band calculation, we take Fourier components within the radius of $|\bm{G}|\leq 2\max\left(|n|,|m|,|n'|,|m'|\right)$ as the basis of Hamiltonian.
We neglect remote interlayer hoppings between layer 1 and 3.
}


    \begin{figure*}
    \begin{center}
    \leavevmode\includegraphics[width=0.9 \hsize]{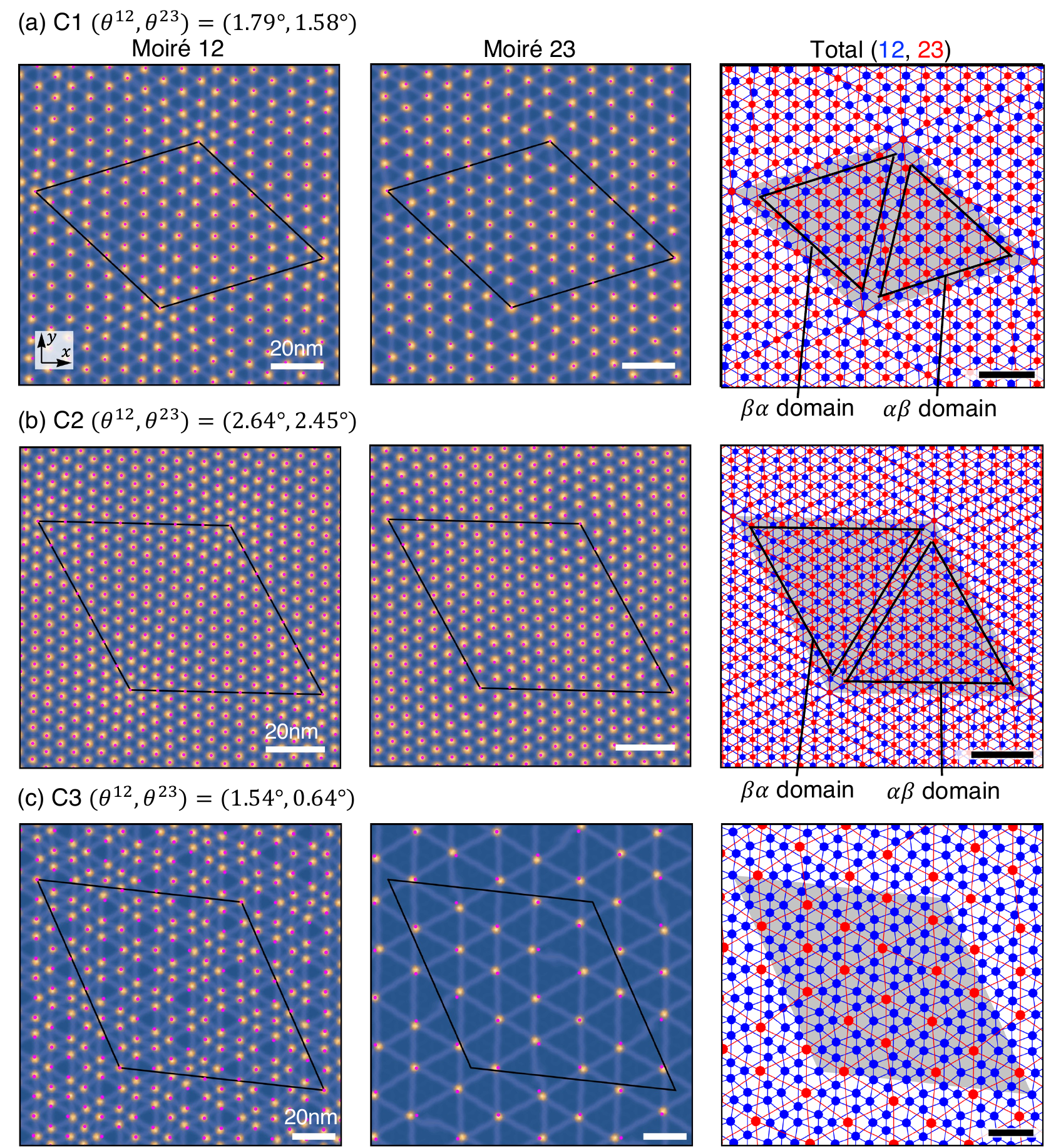}
    \caption{
    Relaxed moir\'e patterns in chiral TTGs,
    (a) C1 $(\theta^{12},\theta^{23})=(1.79^{\circ},1.58^\circ)$, (b) C2 $(2.64^{\circ},2.45^\circ)$ and
    (c) C3 $(1.54^{\circ},0.64^\circ)$.
    In the each row, the left and middle panels are the moiré 12 (between layer 1 and 2) and moiré 23 (between layer 2 and 3) patterns after the relaxation. The color corresponds the local interlayer binding energy $U^{ll'}_{B}$, where bright and dark regions correspond to the AA stack and AB/BA stack respectively.
    Small magenta dots indicate the AA stack points without lattice relaxation for the reference.
    The right panel combines the two moiré patterns in a single plot, where blue and red points indicate the AA stack of the moir\'e 12 and 23 respectively.
    Black triangles represent $\alpha\beta/\beta\alpha$ domains.
     A rhombus in each panel shows the moir\'e-of-moir\'e unit cell and all scale bars incidate 20 nm.
    }
    \label{Fig5}
    \end{center}
    \end{figure*}

\section{Chiral TTGs}
\label{sec_chiral}

\subsection{Multi-scale lattice relaxation}
\label{sec_lattice_relaxation_chiral}

{
We  study the lattice relaxation
in the TTGs of C1$(1.79^\circ,1.58^\circ)$, 
C2$(2.64^\circ,2.45^\circ)$ and C3$(1.54^\circ,0.64^\circ)$ by using the method described in Sec.~\ref{sec:mslr}.
Figure \ref{Fig5} summarizes the optimized moir\'e structures for the three systems.
In each row, the left panel shows
the moir\'e pattern 12 (given by layer 1 and 2),
and the middle panel shows  moir\'e pattern 23 (by layer 2 and 3) after the relaxation.
Here the color represents the local interlayer binding energy $U_B^{ll'}$, where bright and dark regions correspond to the AA stack and AB/BA stack respectively.
Tiny magenta dots indicate the original AA stack points without lattice relaxation for reference.
In the right-most panel, we overlap
the two moir\'e structures in a single diagram, where blue and red points represent the AA stack of the moir\'e 12 and 23 respectively.
A rhombus in each panel represents the moir\'e-of-moir\'e unit cell, and all scale bars indicate 20 nm. 
}

{
We first consider C1 and C2 which have
relatively close twist angles $(\theta^{12},\theta^{23})$.
In the rightmost panels of Fig.~\ref{Fig5} (a) and (b), we see that locally-commensurate $\alpha\beta$ and $\beta\alpha$ domains (indicated by triangles) are formed. 
In these domains,
the lattice relaxation equalizes the two moir\'e periods which were initially different,
to achieve a commensurate structure.
At the same time, we also have the lattice relaxation 
in a smaller scale as in twisted bilayer graphene, which shrinks AA regions and expands AB/BA regions in each of two moir\'e patterns.
Therefore we have the relaxations in 
the moir\'e-of-moir\'e scale 
($\alpha\beta$/$\beta\alpha$ domains)
and in moir\'e scale (AB/BA domains) at the same time.
The following questions naturally arise: (i) What distribution of displacement vectors leads to the multi-scale lattice relaxation? and (ii) Why does such a structure exhibit energetic preference?
These questions can be answered by examining the obtained lattice displacement as follows.
}



    \begin{figure*}
    \begin{center}
    \leavevmode\includegraphics[width=1. \hsize]{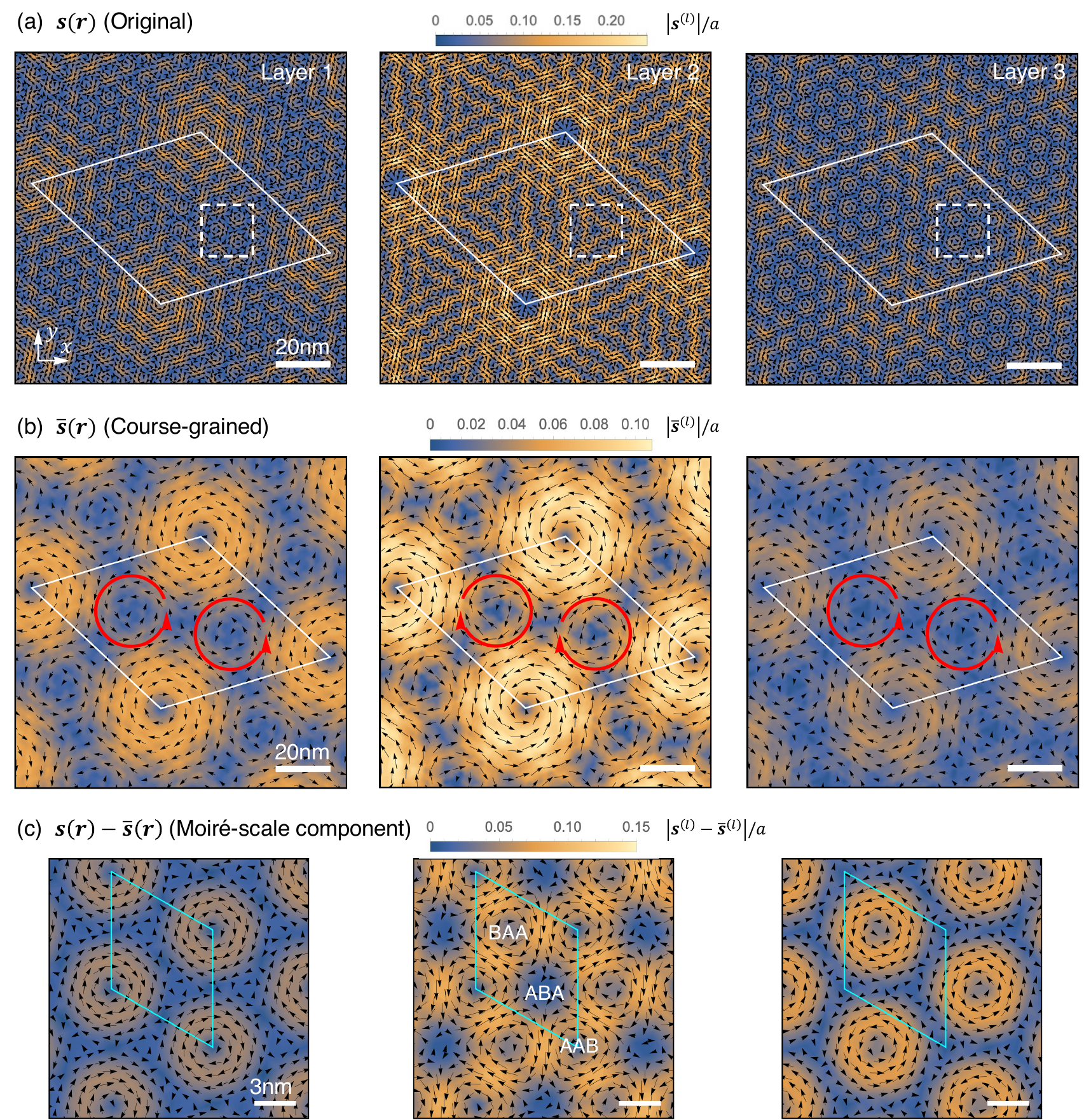}
    \caption{
    Distribution of the displacement vector in each layer of C1:$(\theta^{12},\theta^{23})=(1.79^{\circ},1.58^\circ)$.
    (a) Original non-averaged distribution $\bm{s}^{(l)}(\bm{r}) (l=1,2,3)$.
    (b) Coarse-grained component $\bar{\bm{s}}^{(l)}(\bm{r})$.
    (c) Moir\'e-scale component $\bm{s}^{(l)}(\bm{r})-\bar{\bm{s}}^{(l)}(\bm{r})$ in a region indicated by the white square in the top panel.
    Black arrows represent the displacement vector, and color indicates its norm. Red arc arrows schematically show the direction of rotation in moir\'e-of-moir\'e scale.
    In (a) and (b), the white rhombus represents a moir\'e-of-moir\'e unit cell, while in (c) the blue rhombus represents a moir\'e unit cell.
    }
    \label{Fig6}
    \end{center}
    \end{figure*}

Figure \ref{Fig6}(a) shows
the distribution of the displacement vector 
$\bm{s}^{(l)}(\bm{r})$ on layer 1, 2 and 3
for the case of C1.
The middle row, Fig.~\ref{Fig6}(b), plots 
a coarse-grained component $\bar{\bm{s}}^{(l)}(\bm{r})$, which is calculated by averaging $\bm{s}^{(l)}(\bm{r})$ over a scale of moir\'e unit cell around the point $\bm{r}$.
The bottom row [Fig.\ref{Fig6}(c)] displays magnified plots of $\bm{s}^{(l)}(\bm{r})-\bar{\bm{s}}^{(l)}(\bm{r})$ (i.e., the local component with the coarse-grained part subtracted) within the region enclosed by a dashed square in Fig.\ref{Fig6}(a).

    \begin{figure}
    \begin{center}
    \leavevmode\includegraphics[width=1. \hsize]{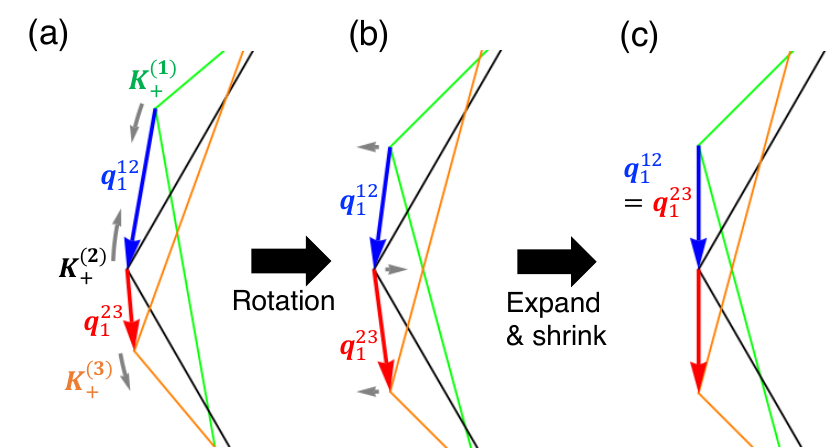}
    \caption{
Relocation of BZ corners 
in the C1 system under the lattice relaxation.
The panels depict: (a) the original non-distorted configuration, (b) the configuration with rotation included, and (c) with expansion and shrinkage taken into account.
Green, black and orange line are the BZ of layer 1, 2 and 3, and gray arrows indicate the direction of rotation and expansion/shrink.
    }
    \label{Fig7}
    \end{center}
    \end{figure}
    
In Fig.~\ref{Fig6}(b), we clearly see that $\bar{\bm{s}}^{(1)}$ and $\bar{\bm{s}}^{(3)}$ rotate counter-clockwise around the center of the $\alpha\beta$ and $\beta\alpha$ domains,
while $\bar{\bm{s}}^{(2)}$ rotates in the clockwise direction.
This behavior is closely linked to
$\alpha\beta/\beta\alpha$ domain formation, and it can be comprehended by examining the problem in the $k$-space.
Figure \ref{Fig7} depicts the relocation of BZ corners of layer 1, 2 and 3 in the C1 system under the lattice relaxation.
The panel (a) is for the original non-distorted configuration.
We define $\bm{q}_{1}^{12}=\bm{K}_{+}^{(2)}-\bm{K}_{+}^{(1)}$ and $\bm{q}_{1}^{23}=\bm{K}_{+}^{(3)}-\bm{K}_{+}^{(2)}$,
where $\bm{K}_{+}^{(l)}$ is the BZ corner of layer $l$ near $\xi=+$ valley.
The vectors $\bm{q}_{1}^{12}$ and $\bm{q}_{1}^{23}$ are associated with the periods of the moir\'e pattern 12 and that of 23, respectively. When these vectors are equal, two moir\'e periods completely match.



The lattice displacement in Fig.~\ref{Fig6}(b)  works precisely to align the two vectors.
In the case of C1, the angle between layer 1 and 2 is larger than the angle between layer 2 and 3 ($\theta^{12}>\theta^{23}$), so the layer 2 rotates clockwise, and the layer 1 and layer 3 rotate counter-clockwise to achieve $\theta^{12}=\theta^{23}$
[Fig.~\ref{Fig7}(b)].
There is still a tiny angle difference between  $\bm{q}_{1}^{12}$ and $\bm{q}_{1}^{23}$.
This can be eliminated by slightly expanding 
BZs layer 1 and 3, and shrinking BZ of layer 2,
to finially obtain the perfect matching [Fig.~\ref{Fig7}(c)].
In the real space, this corresponds to a shrink
of layer 1 and 3 and an expansion of layer 2.
These changes are actually observed in Fig.~\ref{Fig6}(a),
where the vector fields rotate around the center of the $\alpha\beta$/$\beta\alpha$ domain.




{
To understand the energetic stability of $\alpha\beta$/$\beta\alpha$ domains, we examine the local moir\'e-scale lattice relaxation.
Let us first consider the twisted bilayer graphene,
which has only a single moir\'e pattern.
There the lattice relaxation takes place 
such that AB/BA stack region expands and AA stack region shrinks \cite{nam2017lattice}.
This is realized by a local interlayer rotation around AA and AB/BA stack points. 
Around AB/BA, specifically, the layer 1 and 2 oppositely rotate to reduce the local twist angle. The AB/BA region is then enlarged, because the length scale of the moir\'e pattern is enlarged in decreasing the twist angle.
In AA spots, on the contrary, 
the layer 1 and 2 rotate to increase the local twist angle to shrink the AA region.
}

{
The same deformation occurs also in TTG, where all three layers undergo relaxation to expand AB/BA domain
in each of the two moir\'e patterns.
However, as the middle layer $l=2$ is shared by the two interference patterns, there can be a frustration such that, for instance, a local movement of the layer 2 leads to the expansion of the AB region in one moiré pattern while causing its contraction in the other.
Therefore, the relative displacement of the two
moir\'e superlattices should be determined in such a way that the middle-layer distortion can lower the total energies of the two moir\'e patterns at the same time.}

    \begin{figure}
    \begin{center}
    \leavevmode\includegraphics[width=0.9 \hsize]{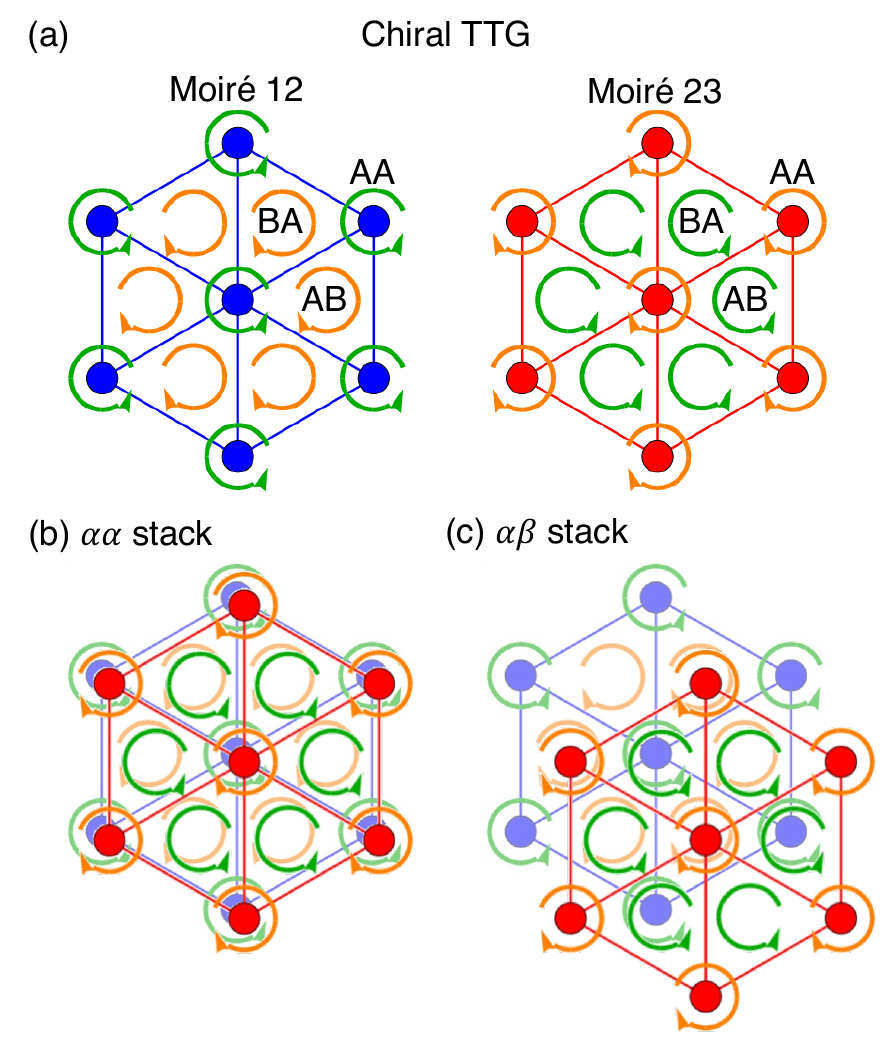}
    \caption{
    (a) Schematic figure of the preferred direction of the middle layer ($l=2)$, for the moir\'e 12 (between $l=1,2$) and moir\'e 23 (between $l=2,3$).
Orange and green arc arrows correspond to clockwise and counterclockwise directions, respectively.
Bottom row: Overlapped figures for (b) $\alpha\alpha$ stack and (c) $\alpha\beta$ stack.
    }
    \label{Fig8}
    \end{center}
    \end{figure}
    
{
Figure \ref{Fig8}(a) is the schematic figure to illustrate the favorable local rotation of the middle layer, for the moir\'e 12 (between $l=1,2$) and moir\'e 23 (between $l=2,3$).
The orange and green arc arrows correspond to clockwise and counterclockwise directions, respectively.
Here we notice that the direction of rotation is opposite for moir\'e 12 and moir\'e 23,
since layer 1 and layer 3 are originally twisted in opposite directions with respect to layer 2.
When AA stack points of moir\'e 12 and  moir\'e 23 are aligned ($\alpha\alpha$ stacking),
the rotation direction of layer 2 is completely frustrated as shown in Fig.~\ref{Fig8}(b), and therefore
$\alpha\alpha$ stacking is energetically unfavorable.
The optimized structure is $\alpha\beta$ stacking
[Fig.~\ref{Fig8}(c)],
where the rotation angles coincide in
two out of three regions.
}


    

{
When the two angles $\theta^{12}$ and $\theta^{23}$ are not close to each other, 
$\alpha\beta$/$\beta\alpha$ domains do not appear  any more, but still a locally-commensurate moir\'e-of-moir\'e structure emerges.
Figure \ref{Fig5}(c) shows the relaxed structure for the C3 TTG.
Since the unit areas of the two moir\'e patterns
differ by nearly 3, we have commensurate domains where a single red triangle includes three blue triangles.
We also see red AA points always come to the center of blue triangles. This can also be understood in terms of the alignment of the favorable rotation angles explained above.
}

    \begin{figure*}
    \begin{center}
    \leavevmode\includegraphics[width=1. \hsize]{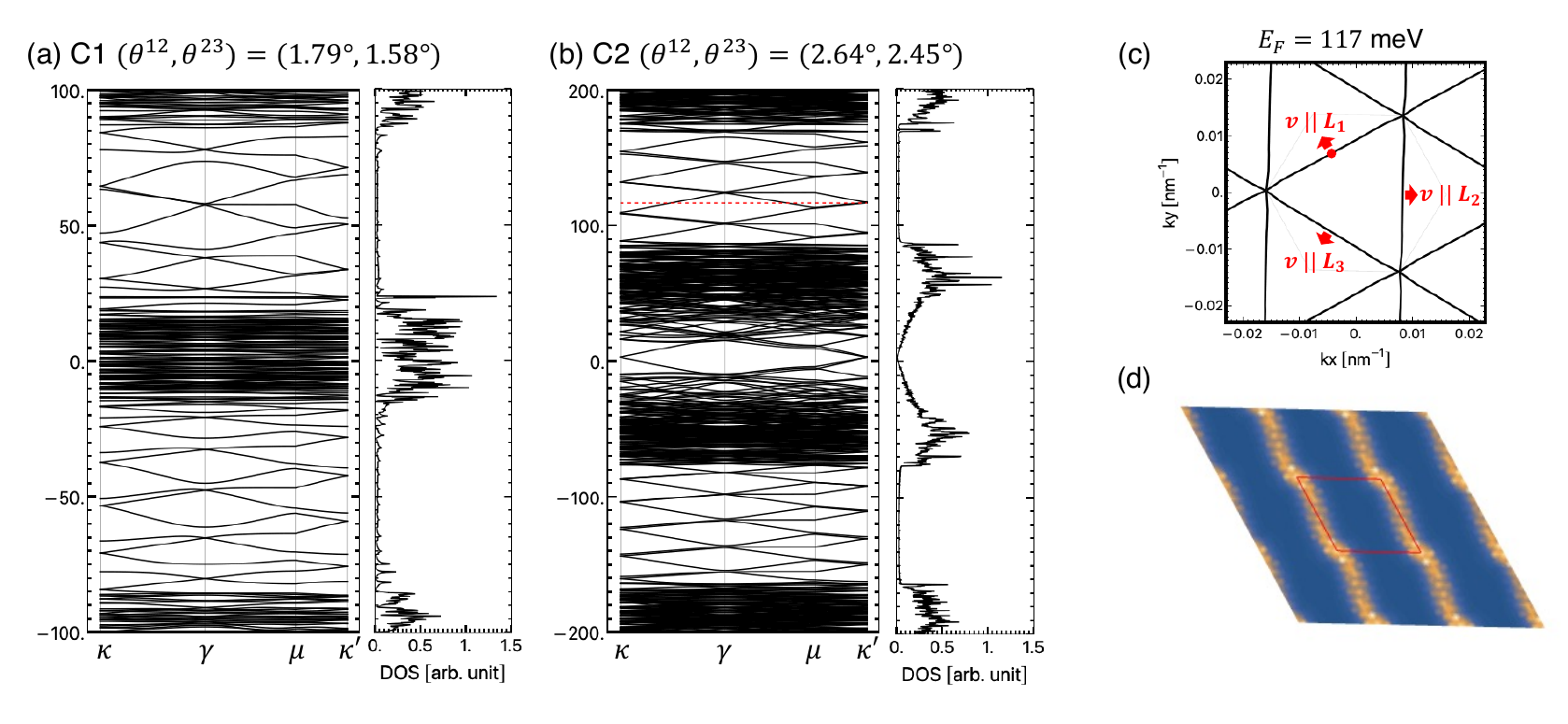}
    \caption{
    (a,b) Electronic band structures and the density of states of $K_{+}$-valley calculated for (a) C1 and (b) C2 with the lattice relaxation incorporated.
    The $k$-space path $(\kappa-\gamma-\mu-\kappa')$ is defined in Fig.~\ref{Fig4}. 
    (c) Fermi surface of the C2 at $E_F=117$ meV (indicated by a red dotted horizontal line in (b)).
    Three red arrows represent the directions of band velocities, which are parallel to the moir\'e lattice vectors $\bm{L}_{1}$, $\bm{L}_{2}$, and $\bm{L}_{3}(=-\bm{L}_{1}+\bm{L}_{2})$.
    (d) Distribution of the squared wave amplitude of an eigenstate state, indicated by a red point in (c).
    Red rhombus represents a moir\'e-of-moir\'e unit cell.
    }
    \label{Fig9}
    \end{center}
    \end{figure*}

\subsection{Electronic properties}
\label{sec_elec_chiral}

{
Using the electronic continuum model introduced in Sec.~\ref{subsec:CM}, we calculate the band structure of TTGs in the presence of the lattice relaxation.
Figure \ref{Fig9}(a) and (b) show the energy bands (near $K_+$ valley) and the corresponding density of states (DOS) calculated for the case C1 and C2, respectively.
The labels $\kappa, \gamma, \mu, \kappa'$ are symmetric points of the moir\'e-of-moir\'e BZ defined in Fig.~\ref{Fig4}.
}

{
We immediately notice that the spectrum
exhibits distinct energy windows characterized by relatively low DOS, which span in the enegy range of 20~meV $< |E| <$ 90~meV for C1, and in 90~meV $< |E| <$ 180~meV for C2.
The windows are sparsely filled with energy bands.
Figure \ref{Fig9}(c) shows the Fermi surface at $E_F=117$ meV in the C2, which is indicated by horizontal red line in Fig.~\ref{Fig9}(b).
We see that the Fermi surface is composed of three intersecting lines arranged with a trigonal symmetry, indicating the dispersion is nearly one-dimensional. The band velocities of these one-dimensional bands (normal to the Fermi surface) are oriented to
the moir\'e-of-moir\'e lattice vectors
 $\bm{L}_1$, $\bm{L}_2$ and $\bm{L}_3(=-\bm{L}_1+\bm{L}_2)$.
Figure \ref{Fig9}(d) plots the distribution of the squared wave amplitudes of an eigenstate marked by a red point in Fig.~\ref{Fig9}(c).
The wave function actually takes a highly one-dimensional form, and it is sharply localized within the domain walls dividing $\alpha\beta$ and $\beta\alpha$ regions.
Each of the three Fermi surfaces corresponds to  one-dimensional states running along the domain walls in the corresponding directions. The states with different directions are barely hybridized.
We also have a low-DOS region near $E=0$ in the C2,
while this is remnant of the graphene's Dirac cone and the energy bands are not one-dimensional.
}


    \begin{figure}
    \begin{center}
    \leavevmode\includegraphics[width=1. \hsize]{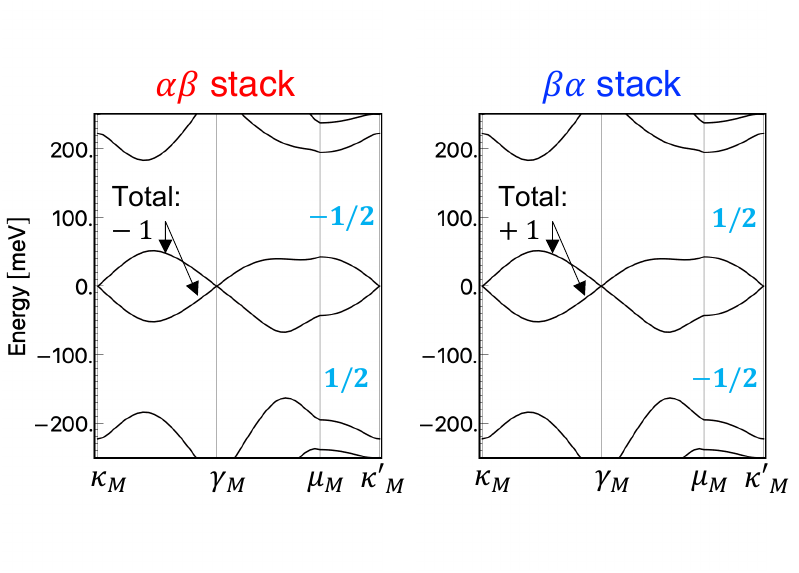}
    \caption{
    Local band structure of the $\alpha\beta$ (left) and $\beta\alpha$ (right) structure with $\theta=2.54^\circ$.
    Black and blue numbers indicate the Chern numbers for 
    bands and gaps, respectively.
    $\kappa_{M}, \gamma_{M}, \mu_{M}, \kappa'_{M}$ are the labels for the common moir\'e BZ, where $\kappa_{M}$ and $\kappa'_{M}$ are corner points, $\mu_{M}$ is the midpoint of a side and $\gamma_{M}$ is the center of the BZ.
    }
    \label{Fig10}
    \end{center}
    \end{figure}

{
The existence of one-dimensional channels on the domain walls indicates that the 
$\alpha\beta$ and $\beta\alpha$ regions are 
locally gapped with different topological numbers,
and associated topological boundary modes emerge between the domains, as shown in Fig.~\ref{Fig0}.
To verify this, we calculate the bands structures and the Chern numbers of {\it uniform} TTG having $\alpha\beta/\beta\alpha$ stacking.
The Hamiltonian of such a uniform system can be obtained by assuming the BZ-corner arrangement in Fig.~\ref{Fig7}(c),
where $\bm{q}_1^{12} = \bm{q}_1^{23} \equiv \bm{q}$.
This corresponds to a TTG where $\theta^{12}=\theta^{23}$ and the layer 2 is slightly expanded in relative to layer 1 and 3.
The two moir\'e periods then become identical,
and we have $\bm{G}_j^{12} = \bm{G}_j^{23} \equiv \bm{G}^{\rm M}_j$ and $\bm{q}= (2\bm{G}^{\rm M}_1+\bm{G}^{\rm M}_2)/3$.
The Hamiltonian for this system is obtained from Eq.~\eqref{eq_Hamiltonian} as
 \begin{align} \label{eq_Hamiltonian_uniform}
    {H}^{(\xi)} =
	\left(
		\begin{array} {ccc}
		    H(\bm{k}+\xi\bm{q}) & U_{21}^{\dagger} & \\
		    U_{21} & H(\bm{k}) & U_{32}^{\dagger}\\
            & U_{32} & H(\bm{k}-\xi\bm{q}),
			\end{array}
		\right). 
    \end{align}
where 
\begin{align} 
& H(\bm{k}) =-\hbar v \bm{k}\cdot \bm{\sigma},\\
&
U_{21}= \sum_{j=1}^{3} U_{j} \e^{\i \xi\bm{\bm{\delta k}}_{j}\cdot\bm{r}},
\quad
U_{32}= \sum_{j=1}^{3} U_{j} \e^{\i \xi\bm{\bm{\delta k}}_{j}\cdot (\bm{r}-\bm{r}_0)} 
\notag\\
&\delta\bm{k}_{1}=\bm{0},\,\,\,
\delta\bm{k}_{2}=\xi\bm{G}^{\rm M}_{1},\,\,\,
\delta\bm{k}_{3}=\xi\left(\bm{G}^{\rm M}_{1}
+\bm{G}^{\rm M}_{2}\right),
 \end{align}
and we neglect the strain-induced vector potentials which does not affect the topological nature argued here.
Here $U_{21}$ and $U_{32}$ differ by the parameter 
$\bm{r}_0$, which specifies the relative displacement between the two moir\'e patterns.
The $\alpha\beta$ and $\beta\alpha$ stackings correspond to 
$\bm{r}_0=\left(\bm{L}_{1}^{M}+\bm{L}_{2}^{M}\right)/3$ and $2\left(\bm{L}_{1}^{M}+\bm{L}_{2}^{M}\right)/3$ respectively,
where $\bm{L}_{j}^{M}$ is the common moir\'e lattice vector given by $\bm{G}_{i}^{M}\cdot\bm{L}_{j}^{M}=2\pi\delta_{ij}$.



{
Here we consider uniform $\alpha\beta$ and $\beta\alpha$ TTGs with $\theta^{12}=\theta^{23}=2.54^\circ$, 
 which approximate the local structures of $\alpha\beta$ and $\beta\alpha$ domains in the C2.
 Figure \ref{Fig10} plots the energy bands in $\xi=+$ valley calculated by Eq.~\eqref{eq_Hamiltonian_uniform}.
 We observe energy gaps in the electron and hole sides in the region $50\,{\rm meV} < |E| < 180\,{\rm meV}$,
 which approximately coincides with the energy window of the C2 [Fig.~\ref{Fig9}(b)].
Between the gaps in the electron and hole sides,
we have two bands touching at the charge neutrality point.
The total Chern number for the two-band cluster is found to be $\mp 1$ for $\alpha\beta$ and $\beta\alpha$, respectively. The absolute Chern number in the upper gap can also be calculated, and it turns out to be 
$\mp 1/2$ for $\alpha\beta$ and $\beta\alpha$, respectively.
This is obtained by opening mass gap (adding asymmetric energies to A and B sublattices in all the graphene layers) to lift the band touching at the Dirac point.
Since the difference of the Chern number of the upper gap between the $\alpha\beta$ and $\beta\alpha$ regions is 1, we have a single edge mode (per a single valley) at the domain boundary. This coincides with the number of the one-dimenisonal modes per a single direction in the moir\'e-of-moir\'e superlattice band Fig.~\ref{Fig9}.
The Chern number of the valley $\xi=-1$ is negative of 
$\xi=+1$ valley due to the time reversal symmetry.
Therefore the TTG is a quantized valley Hall insulator when the Fermi energy is in the energy window. 
}

{
The energy windows and one-dimensional domain-wall states also appear in the C1 case [Fig.~\ref{Fig9}(a)],
which has a smaller moir\'e-of-moir\'e period.
The degree of one-dimensionality is not as pronounced as in the C2 configuration, as evidenced by the appearance of small gaps at the intersections of bands.
The hybridization tends to be greater when the moir\'e-of-moir\'e period is smaller.
}

{
Finally, the band structure in Fig.~\ref{Fig9}
closely resembles the marginally-stacked twisted bilayer graphene in a strong perpendicular electric field \cite{PhysRevB.88.121408,PhysRevLett.121.146801,PhysRevB.98.035404,fleischmann2019perfect,walet2019emergence,PhysRevB.101.125409,PhysRevB.101.201403}.
There the topological one dimensional edge states arise 
since the AB and BA regions in the moir\'e pattern have opposite valley Chern numbers in the electric field.
The chiral TTG realizes a similar situation in 
the moir\'e-of-moir\'e scale, without the need for 
an applied electric field.
This can be achieved in any chiral TTGs where $\theta^{12}$ and $\theta^{23}$ are close to each other, such that the two moir\'e periods are comparable.
}

    \begin{figure*}
    \begin{center}
    \leavevmode\includegraphics[width=0.9 \hsize]{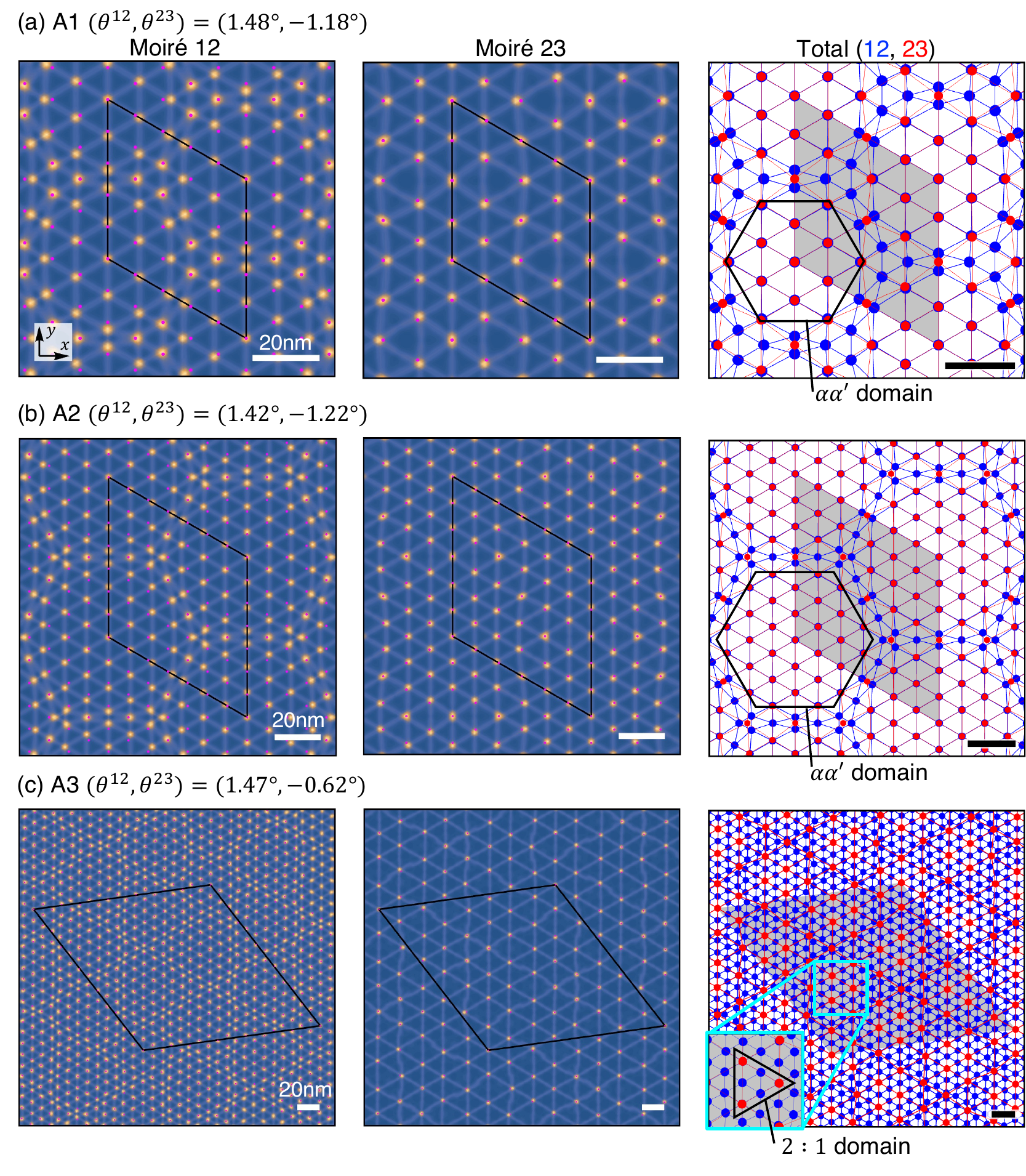}
    \caption{
Relaxed moir\'e patterns in alternating TTGs,  
    (a) A1:$(\theta^{12},\theta^{23})=(1.48^\circ,-1.18^\circ)$, (b) A2: $(1.42^\circ,-1.22^\circ)$ and
(c) A3:$(1.47^\circ,-0.62^\circ)$,
corresponding to Fig. \ref{Fig5} for the chiral TTGs.
    }
    \label{Fig11}
    \end{center}
    \end{figure*}

    \begin{figure*}
    \begin{center}
    \leavevmode\includegraphics[width=1. \hsize]{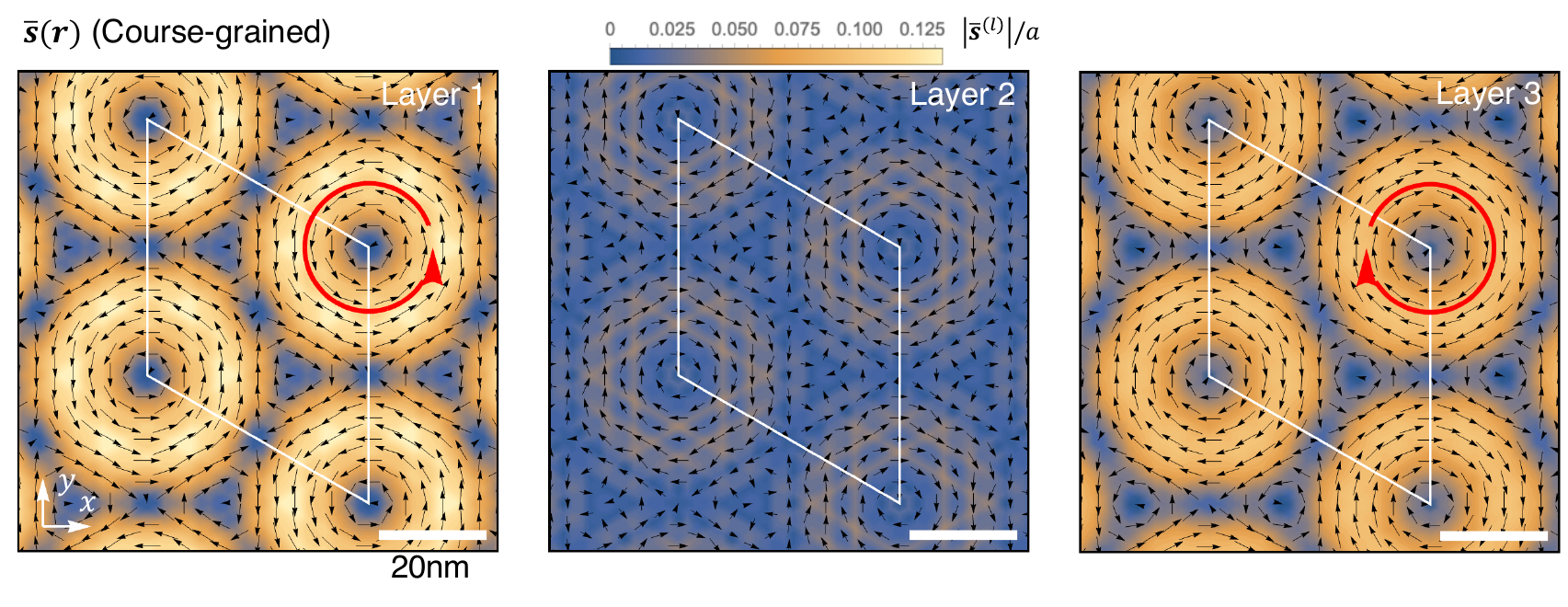}
    \caption{
    Distribution of the coarse-grained displacement vector $\bar{\bm{s}}^{(l)}(\bm{r})$ in  A1:$(\theta^{12},\theta^{23})=(1.48^\circ,-1.18^\circ)$, corresponding  to Fig. \ref{Fig6}(b) for the C1.
    }
    \label{Fig12}
    \end{center}
    \end{figure*}

\section{alternating TTGs}
\label{sec_alternating}

\subsection{Multi-scale lattice relaxation}
\label{sec_lattice_relaxation_alternating}

{
Alternating TTGs display distinct relaxed structures that differ entirely from the chiral cases.
Figure \ref{Fig11} shows optimized moir\'e structures 
calculated for alternating TTGs (a) A1 $(\theta^{12},\theta^{23})=(1.48^\circ,-1.18^\circ)$, (b) A2 $(1.42^\circ,-1.22^\circ)$ and
(c) A3 $(1.47^\circ,-0.62^\circ)$,
corresponding to Fig.~\ref{Fig5} for chiral TTGs.
In the A1 and A2, we observe a formation of commensurate $\alpha\alpha'$ domains,
where AA spots of the two moir\'e patterns completely overlaps [See Fig.~\ref{Fig1}(e)].
This is in a sharp contrast to the chiral TTGs, where AA spots are repelled to each other, giving rise to $\alpha\beta/\beta\alpha$ domains.
The atomic structure of $\alpha\alpha'$ domain corresponds precisely to the mirror-symmetric TTG with $\theta^{12} = -\theta^{23}$.
In A3 case [Fig.~\ref{Fig11}(c)], where the two moir\'e periods are not comparable, 
we observe a different type of commensurate domain with the ratio of the lattice periods fixed at 2, 
reflecting the original moir\'e-period ratio $L^{23}/L^{12} \simeq 2.3$.
Here the AA stacking points of the red and blue moir\'e lattices are vertically aligned as in $\alpha\alpha'$ domains observed in A1 and A2}.
}



The formation of the commensurate domains can be attributed to a specific type of lattice distortion that differs from the chiral case.
Figure \ref{Fig12} shows the distribution of the coarse-grained displacement vector 
$\bar{\bm{s}}^{(l)}(\bm{r})$ in the A1 case (corresponding to Fig.~\ref{Fig6}(b) for the chiral case).
We observe that the layer 1 and layer 3 rotate anti-clockwise and clock-wise directions, respectively, around $\alpha\alpha'$ domain center.
{
In $k$-space, accordingly, the Brillouin zone corners of layer 1 and 3 move to overlap as shown in Fig.~\ref{Fig_moire_match_alternating}. This corresponds to the symmetric TTG $(\theta^{12} = -\theta^{23})$ where the layer 1 and layer 3 are perfectly aligned.
}

    \begin{figure}
    \begin{center}
    \leavevmode\includegraphics[width=1. \hsize]{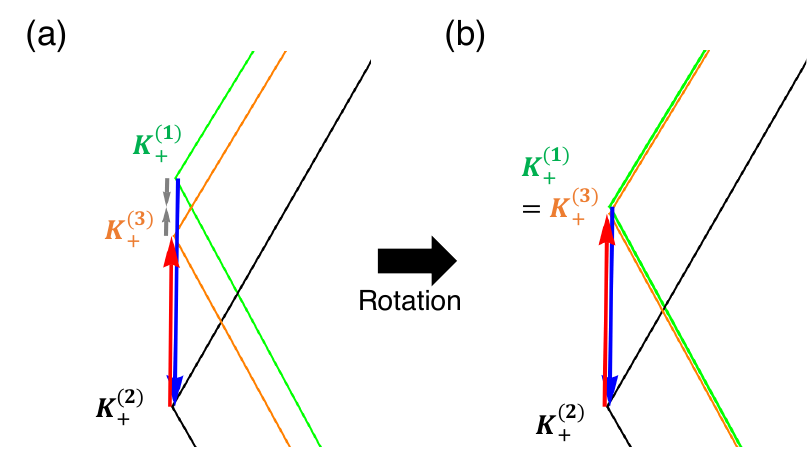}
    \caption{
Relocation of BZ corners 
in the A1:$(\theta^{12},\theta^{23})=(1.48^\circ,-1.18^\circ)$ under the lattice relaxation.
The panels depict (a) the original non-distorted configuration and (b) the relaxed configuration.
    }
    \label{Fig_moire_match_alternating}
    \end{center}
    \end{figure}

{
The stability of $\alpha\alpha'$-domain is also explained by considering moir\'e-scale lattice relaxation.
As discussed in Sec. \ref{sec_lattice_relaxation_chiral}, 
the graphene layers in TTG undergo spontaneous distortion to expand the AB/BA regions for the moir\'e patterns 12 and 23, giving a competitive environment for the shared layer 2.
Figure \ref{Fig13}(a) depicts the preferred orientation of layer 2 for the two moir\'e patterns in alternating TTG.
In contrast to the chiral stack [Fig. \ref{Fig8}],
the rotation direction is identical for both moir\'e patterns, since layer 1 and layer 3 are rotated in the same direction relative to the layer 2.
Consequently, there is no frustration when the moir\'e lattices are arranged in an $\alpha\alpha'$ stack as shown in Fig.~\ref{Fig13}(b). 
In this structure, the motion of the shared layer 2 allows for the simultaneous relaxation of the moir\'e patterns 12 and 23, resulting in an energy advantage compared to partially frustrated configurations like the $\alpha\beta'$ stack [Fig.~\ref{Fig13}(c)].
The stability of $\alpha\alpha'$ stack in nearly-symmetric TTGs was pointed out in the previous theoretical works \cite{doi:10.1021/acs.nanolett.9b04979,PhysRevB.104.035139,PhysRevB.106.075423,meng2023commensurate}}, and it was observed in recent experiments \cite{doi:10.1126/science.abk1895,craig2023local}.

    \begin{figure}
    \begin{center}
    \leavevmode\includegraphics[width=0.9 \hsize]{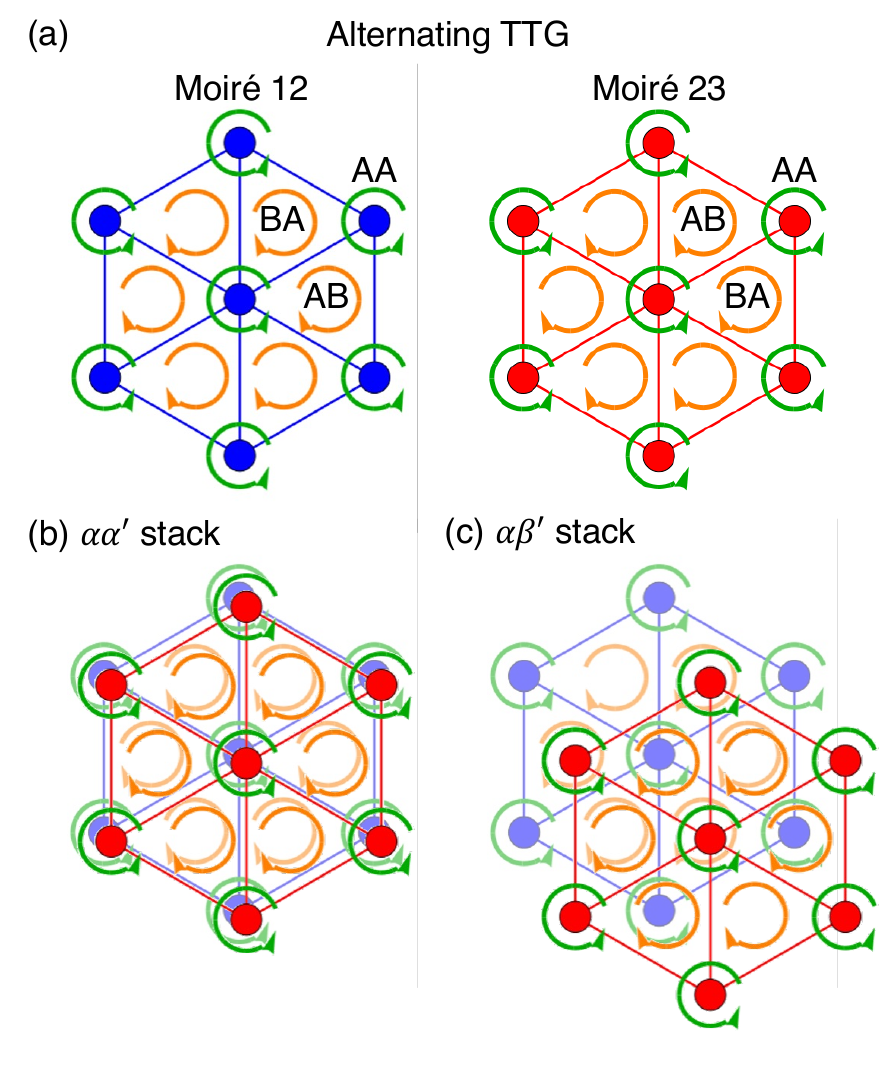}
    \caption{
    (a) Schematic figure of the preferred distorting direction of the middle layer ($l=2)$ in an alternating TTG, corresponding to Fig. \ref{Fig8} for a chiral TTG.
    }
    \label{Fig13}
    \end{center}
    \end{figure}
    \begin{figure*}
    \begin{center}
    \leavevmode\includegraphics[width=1. \hsize]{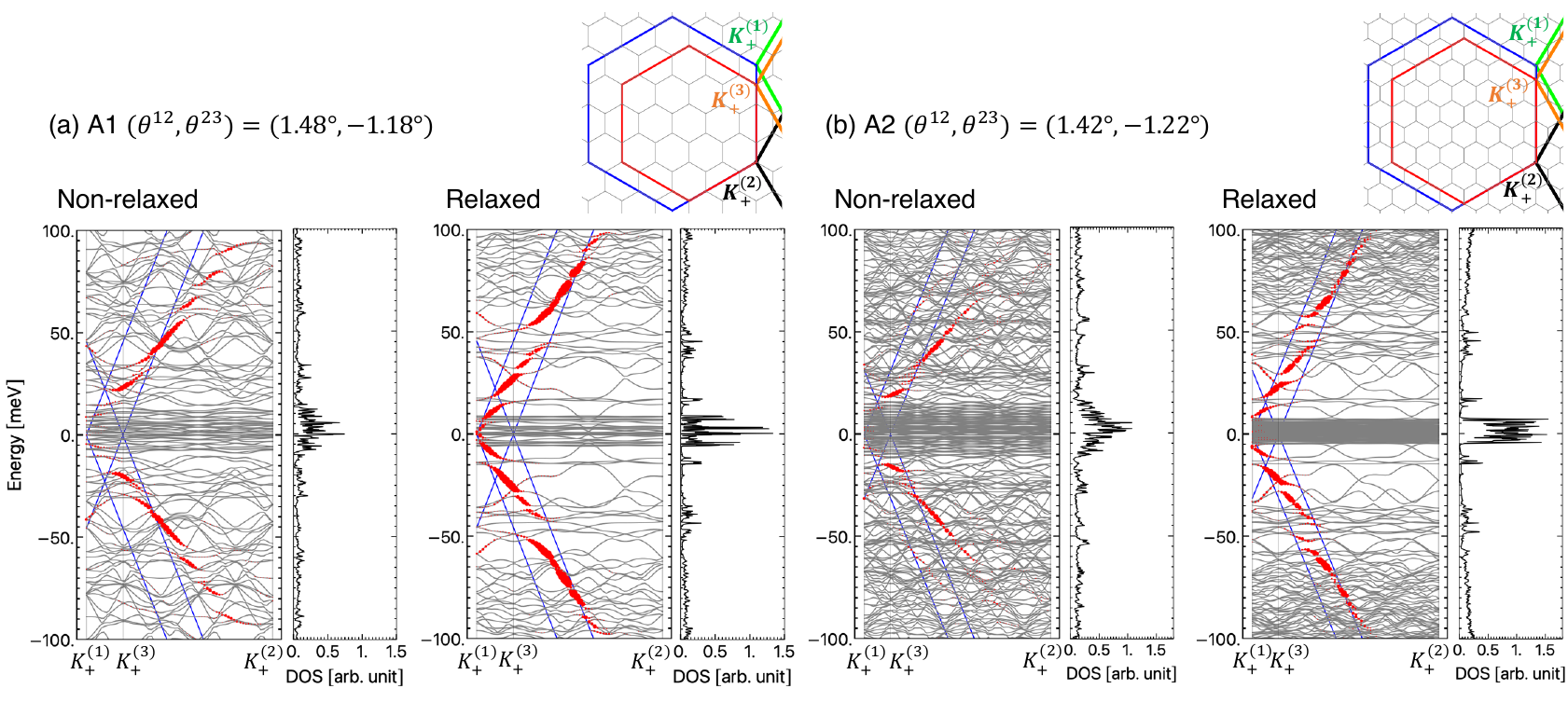}
    \caption{
    Energy bands and DOS for alternating TTGs, (a) A1:$(1.48^{\circ},1.18^\circ)$ and (b) A2:$(1.42^{\circ},1.22^\circ)$.
    The left and right panels in each figure show the results without and with the lattice relaxation, respectively.
    Black curves represent the energy bands,
and blue straight lines indicate the intrinsic Dirac bands of layer 1 and layer 3 without the interlayer coupling.
Red dots indicate the amplitude projected onto the mirror-odd plane wave states (see the text).
The path is taken as $\bm{K}_{+}^{(1)} \to \bm{K}_{+}^{(3)} \to \bm{K}_{+}^{(2)}$
in the extended $k$-space shown in the inset. 
}
    \label{Fig14}
    \end{center}
    \end{figure*}

\subsection{Electronic properties}
\label{sec_elec_alternating}

{
We calculate the band structure for alternating TTGs of A1$(1.48^{\circ},-1.18^\circ)$, (b)A2$(1.42^{\circ},-1.22^\circ)$
using the method described in Sec.~\ref{sec_model}.
The energy band and DOS for A1 and A2
are displayed in Figs.~\ref{Fig14}(a) and (b), respectively.
In each figure, the right and left panels correspond to the TTGs with and without the lattice relaxation, respectively.
Black curves represent the energy bands,
and blue straight lines indicate the intrinsic Dirac bands of layer 1 and layer 3 without the interlayer coupling.
Red dots indicate the amplitude projected onto the mirror-odd plane wave states, as defined by
\begin{align} \label{eq:w_odd}
   & w_{n\bm{k}}^{\rm (odd)}
   = \sum_{X=A,B}
   |\langle \psi_{n\bm{k}}|
   \bm{k},X,{\rm odd}\rangle|^2 ,
   \notag\\
&|\bm{k},X,{\rm odd}\rangle   
 = \frac{1}{\sqrt{2}}
\left(
|\bm{k},X,1\rangle -|\bm{k},X,3\rangle
\right),
\end{align}
where $\psi_{n\bm{k}}$ is the eigenstates, and $|\bm{k},X,l\rangle$ is the plane wave at sublattice $X (=A,B)$ on layer $l$.
We take the path $\bm{K}_{+}^{(1)} \to \bm{K}_{+}^{(3)} \to \bm{K}_{+}^{(2)}$ 
on a straight line in the extended $k$-space, as shown in insets of
Fig.~\ref{Fig14}.
}



{
In the band structures with the lattice relaxation, we observe numerous flat bands concentrated around zero energy, and these bands are surrounded by a region where dispersive energy bands are sparsely distributed.
These features coincide with the mirror-symmetric TTG $(\theta^{12}=-\theta^{23})$,
where the low-energy spectrum is composed of a flat band with even parity, and a Dirac cone with odd parity against the mirror inversion \cite{li2019electronic,doi:10.1021/acs.nanolett.9b04979}. 
We see that the red dots roughly form a conical dispersion, and it is regarded as a remnant of the symmetric TTG's Dirac cone having odd parity.
In the non-relaxed calculations, we notice that the flat bands and Dirac cones are strongly hybridized, and the conical dispersion of the red dots is not clearly resolved. These results suggest that the formation of $\alpha\alpha'$ domains (equivalent to the mirror-symmetric TTG) supports the spectral separation of the flat bands and the Dirac-cone like bands. Therefore, we expect that asymmetric TTGs slightly away from the symmetric condition $\theta^{12}=-\theta^{23}$ acquire similar electronic properties to the symmetric TTG, through the moir\'e-of-moir\'e lattice relaxation.
}

{
The electronic properties of TTG can be tuned by applying a perpendicular electric field.
We can introduce the field effect to our model as $H+V$, where $H$ is the original Hamiltonian of Eq.~\eqref{eq_Hamiltonian},
and $V$ is the on-site potential term by perpendicular electronic field,
 \begin{align} \label{eq_u_mat}
	V=\left(
		\begin{array}{ccc}
		  -\Delta \hat{I}_{2} & & \\
		   & 0 &  \\
             & & \Delta \hat{I}_{2} 
		\end{array}
	\right).
    \end{align}
Here $\Delta$ is the difference of the on-site energy and $\hat{I}_{2}$ is a $2\times2$ unit matrix, and we simply assumed the perpendicular electric field is constant between top layer and bottom layer.
Figure \ref{Fig15} shows the energy band of the A2 with lattice relaxation, under the perpendicular electric field $\Delta = 50$ meV and $100$ meV.
When the electric field is applied, we observe the Dirac band moves along the energy axes, and eventually the Dirac point emerges out of the flat-band cluster.
We also see that the electric fields broadens the energy width of the flat band region, and enhances a hybridization between the flat bands and the dispersive bands.
}



    \begin{figure}
    \begin{center}
    \leavevmode\includegraphics[width=1. \hsize]{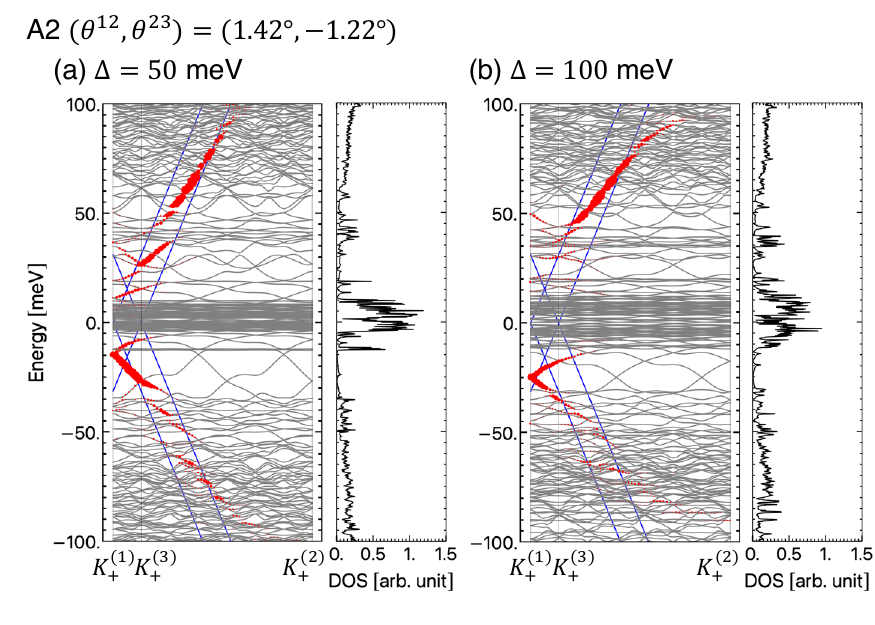}
    \caption{
    Plots similar to Fig.~\ref{Fig7} for A2:$(1.43^{\circ},-1.28^\circ)$ with the perpendicular electric field of (a) $\Delta=50$ meV and (b) $100$ meV.
    }
    \label{Fig15}
    \end{center}
    \end{figure}

\section{Conclusion}
\label{sec:con}

{
We have presented a systematic investigation on the lattice relaxation and electronic properties of general non-symmetric TTGs.
For various chiral and alternating TTGs with different twist angle combinations, we employ an effective continuum approach similar to twisted bilayer graphene, to obtain the optimized lattice structure.
We also computed the electronic band structure by using a continuum band calculation method incorporating lattice relaxation effects.
In the calculation of the lattice relaxation, we found that there are two distinct length-scale relaxations in the moir\'e-of-moir\'e and moir\'e scales, which lead to the formation of a patchwork of super-moiré domains.
In these domains, two moiré patterns become locally commensurate with a specific relative arrangement.
The chiral TTGs prefer a shifted stacking
where the overlap of AA spots in the individual moir\'e patterns is avoided.
In contrast, the alternating TTGs exhibits a completely opposite behavior where AA spots are perfectly overlapped.
In the band calculations,
the chiral TTG exhibits an energy window where highly one-dimensional electron bands are sparsely distributed. By calculating the Chern number of the local band structure within the commensurate domains, we identify one-dimensional domain boundary states
as topological boundary states between distinct Chern insulators.
The alternating TTG exhibits a clear separation of the flat bands and a monolayer-like Dirac cone, as a consequence of the formation of commensurate domains equivalent to the symmetric TTG.
}
%
%

{\it Note added}: During the finalization of this paper, we became aware of related preprints which partially overlap with the present work
\cite{devakul2023magic,guerci2023chern}.

\section*{Acknowledgments}

This work was supported in part by 
JSPS KAKENHI Grant Number JP20H01840, JP20H00127, JP21H05236, JP21H05232 and by JST CREST Grant Number JPMJCR20T3, Japan.


\bibliography{reference}
\end{document}